\documentclass[prx,aps,amsmath,amssymb,showpacs,twocolumn]{revtex4-1}
\usepackage{amssymb,graphicx,float,dcolumn,bm,subfigure,color}
\usepackage[titletoc,title]{appendix}
\usepackage[dotinlabels]{titletoc}
\usepackage{cancel}
\usepackage{subfigure}
\usepackage{braket}
\usepackage{mathrsfs}
\usepackage[mathscr]{euscript}
\usepackage{hyperref}

\begin{document}

\title{$\mathbb{Z}_{n}$ superconductivity of composite bosons and the $7/3$ fractional quantum Hall effect}
\author{Ajit C. Balram$^{1,2}$, J. K. Jain$^{3}$ and Maissam Barkeshli$^{4}$}
\affiliation{$^{1}$The Institute of Mathematical Sciences, HBNI, CIT Campus, Chennai 600113, India}
\affiliation{$^{2}$Niels Bohr International Academy and the Center for Quantum Devices, Niels Bohr Institute, University of Copenhagen, 2100 Copenhagen, Denmark}
\affiliation{$^{3}$Department of Physics, 104 Davey Lab, Pennsylvania State University, University Park, Pennsylvania 16802, USA}
\affiliation{$^{4}$Condensed Matter Theory Center and Joint Quantum Institute, Department of Physics, University of Maryland, College Park, Maryland 20472 USA}
\date{\today}

\begin{abstract} 
The topological $p$-wave pairing of composite fermions, believed to be responsible for the 5/2 fractional quantum Hall effect (FQHE), has generated much exciting  physics. Motivated by the parton theory of the FQHE, we consider the possibility of a new kind of emergent ``superconductivity" in the 1/3 FQHE, which involves condensation of clusters of $n$ composite bosons. From a microscopic perspective, the state is described by the $n\bar{n}111$ parton wave function ${\cal P}_{\rm LLL} \Phi_n\Phi_n^*\Phi_1^3$, where $\Phi_n$ is the wave function of the integer quantum Hall state with $n$ filled Landau levels and ${\cal P}_{\rm LLL}$ is the lowest-Landau-level projection operator. It represents a $\mathbb{Z}_{n}$ superconductor of composite bosons, because the factor $\Phi_1^3\sim \prod_{j<k}(z_j-z_k)^3$, where $z_j=x_j-iy_j$ is the coordinate of the $j$th electron, binds three vortices to electrons to convert them into composite bosons, which then condense into the $\mathbb{Z}_{n}$ superconducting state $|\Phi_n|^2$. From a field theoretical perspective, this state can be understood by starting with the usual Laughlin theory and gauging a $\mathbb{Z}_n$ subgroup of the $U(1)$ charge conservation symmetry.  We find from detailed quantitative calculations that the $2\bar{2}111$ and $3\bar{3}111$ states are at least as plausible as the Laughlin wave function for the exact Coulomb ground state at filling $\nu=7/3$, suggesting that this physics is possibly relevant for the 7/3 FQHE.  The $\mathbb{Z}_{n}$ order leads to several observable consequences, including quasiparticles with fractionally quantized charges of magnitude $e/(3n)$ and the existence of multiple neutral collective modes. It is interesting that the FQHE may be a promising venue for the realization of exotic $\mathbb{Z}_{n}$ superconductivity. 
\end{abstract}

\maketitle

\section{Introduction}
\label{sec:introduction}

The system of interacting electrons in the lowest Landau level (LLL) [or, in general, a given LL] has given rise to some of the most profound emergent structures found in nature. The first clue came from the observation of  fractional quantum Hall effect (FQHE) at filling factor $\nu=1/3$~\cite{Tsui82} in the lowest Landau level (LLL). This was explained by Laughlin through construction of an ansatz wave function~\cite{Laughlin83}. Zhang, Hansson and Kivelson~\cite{Zhang89} constructed a Chern-Simons theory of the Laughlin state as a condensate of composite bosons, which are bound states of electrons and an odd number of vortices. In subsequent years a large number of additional states were observed, primarily along the sequences $\nu=n/(2pn\pm 1)$ ($n$ and $p$ are positive integers) and their hole conjugates $\nu=1-n/(2pn\pm 1)$. These are understood as the integer quantum Hall effect (IQHE) of composite fermions (CFs)~\cite{Jain89,Jain07}, which are bound states of electrons and an even number ($2p$) of quantized vortices. Further structure emerges from a chiral $p$-wave pairing of composite fermions, described by the Moore-Read Pfaffian wave function, which produces incompressibility at even denominator fractions~\cite{Moore91,Read00}. This physics is believed to be responsible for FQHE at $\nu=5/2$~\cite{Willett87}, and has generated many remarkable new ideas, such as non-Abelian quasiparticles with localized Majorana zero modes. 

The aim of our present article is to show that it is possible to construct FQHE states that involve $\mathbb{Z}_{n}$ superconductivity in which clusters of $n$ composite bosons condense. (It is stressed that we are dealing here with the FQHE of electrons; the objects that condense are emergent topologically non-trivial bosonic excitations.) We further demonstrate, based on extensive quantitative calculations, that such states are at least as plausible as the Laughlin state in the second LL (SLL), and thus possibly relevant for the 7/3 FQHE. This not only reveals the possibility of a new kind of topological order in FQHE, but also shows that the FQHE may be an ideal place for realizing exotic $\mathbb{Z}_{n}$ superconductivity. 

There are two motivations for this work, a theoretical and an experimental. We begin by discussing them separately.

A large class of candidate FQHE states were proposed by the the parton construction~\cite{Jain89b}. Here, one imagines dividing each electron into $m$ fictitious particles called ``partons," placing each species of partons in an IQH state with filling $n_{\alpha}$ (here $\alpha=1, \cdots, m$ labels the different parton species), and finally gluing the partons back together to recover the physical electrons. The resulting wave function, referred to as the $n_1 n_2 \cdots n_m$ state, is given by
\begin{equation}
\Psi_{\nu}^{n_1 n_2\cdots n_m} = \mathcal{P}_{\rm LLL}\prod_{\alpha=1}^m \Phi_{n_{\alpha}}(\{z_{i}\}).
\label{eq:parton_general} 
\end{equation}
Here we represent the coordinates of the $j$th particle as $z_j=x_j-iy_j$, $\Phi_n$ is the Slater determinant wave function of the IQH state with $n$ filled LLs of electrons, and $\mathcal{P}_{\rm LLL}$ projects the state into the LLL. We will denote a negative integer as $\bar{n}=-n$ with $\Phi_{\bar{n}}\equiv \Phi_{-n}=[\Phi_n]^*$. Note that each of the constituent IQH states is itself made up of \emph{all} of the electrons. The wave function given in Eq.~(\ref{eq:parton_general}) occurs at the filling factor $\nu=\left[\sum_\alpha n_\alpha^{-1}\right]^{-1}$ and has a shift $\mathcal{S}=\sum_\alpha n_\alpha$. (The shift is a topological quantum number which accounts for the offset in the relation between the number of electrons $N$ and number of flux quanta $2Q$ at which an incompressible quantum Hall state occurs in the spherical geometry~\cite{Haldane83}, through the relation $\mathcal{S}=\nu^{-1}N-2Q$~\cite{Wen92}.) The charge of the $\alpha$ parton species is given by $e_{\alpha}=-\nu e/n_\alpha$, which is consistent with the constraint that charges of the partons add to charge of the electron, i.e., $\sum_\alpha e_\alpha=-e$, where $-e$ is the charge of the electron. A field theoretical description of these states has been developed by Wen and collaborators~\cite{Blok90,Blok90b,Wen91,Wen92b}.

The parton theory contains many of the known FQHE states. The Laughlin state appears as $111\cdots$, and the $\nu=n/(2pn+1)$ and $\nu=n/(2pn-1)$ Jain CF states as $n11\cdots$ and $\bar{n}11\cdots$, respectively. Recently, Balram, Barkeshli and Rudner have shown~\cite{Balram18} that the $\bar{2}\bar{2}111$ wave function describes a state in the same universality class as the hole-conjugate of the paired Moore-Read state. Balram and collaborators have further proposed that some other states of the parton theory describe the observed SLL states at $\nu=2+6/13$ ~\cite{Balram18a} and also at $\nu=2+2/5$~\cite{Balram19}. It has been proposed that the $221$ and $22111$ states are applicable to certain even-denominator FQHE states in graphene~\cite{Wu16,Kim18} and wide quantum wells~\cite{Faugno19} respectively.

\begin{figure}[t]
\begin{center}
\includegraphics[width=0.43\textwidth,height=0.37\textwidth]{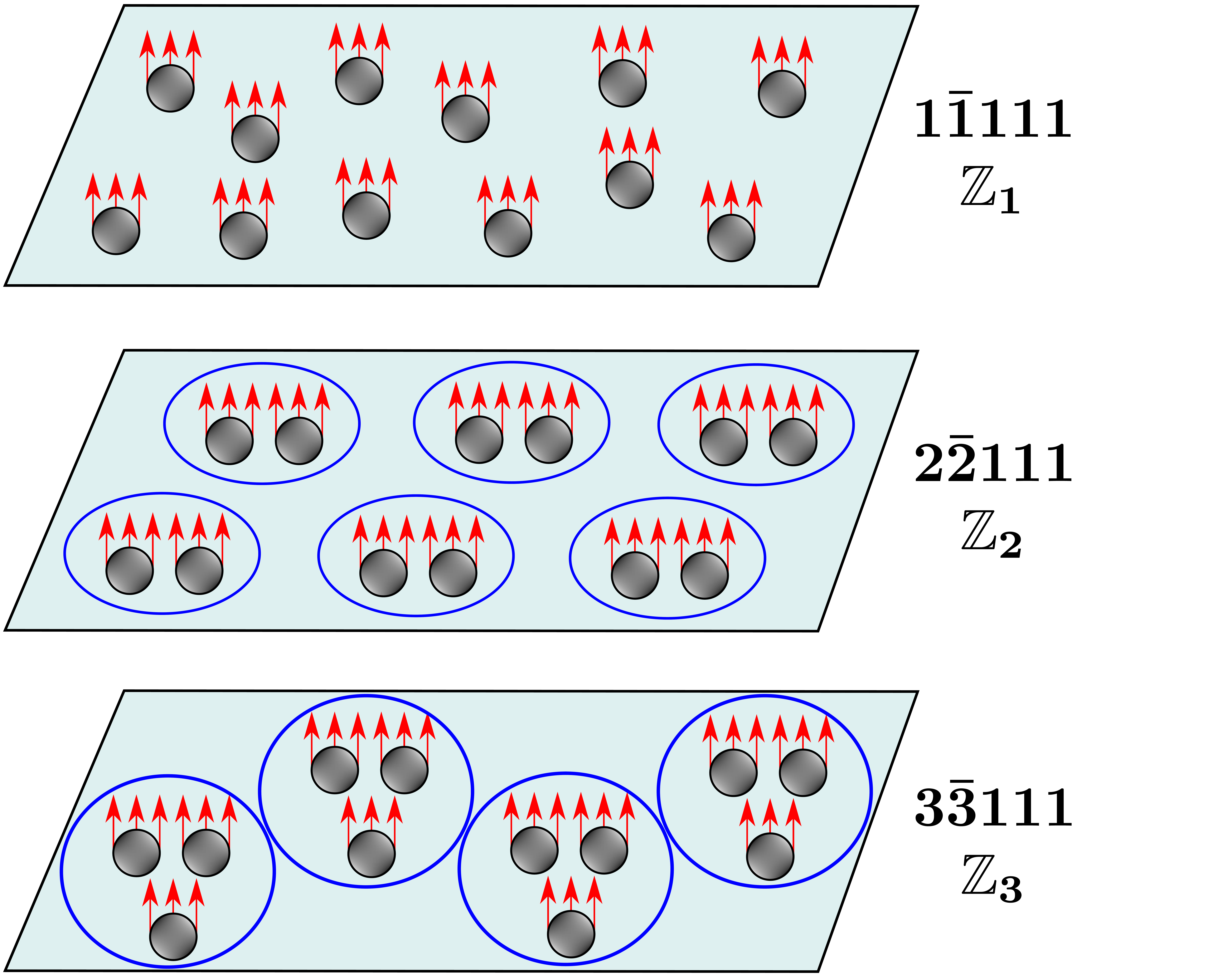} 
\caption{(color online) Schematic illustration that the $n\bar{n}111$ parton state can be interpreted as a $\mathbb{Z}_n$ superconducting state of composite bosons (electron bound to three quantized vortices), wherein clusters of $n$ composite bosons condense.}
\label{fig:schematic}
\end{center}
\end{figure}

The states of our interest in this article are labeled $n\bar{n}111$ and their wave functions are given by
\begin{equation}
\Psi^{n\bar{n}111}_{1/3} = \mathcal{P}_{\rm LLL} \Phi_{n}\Phi_{\bar{n}}\Phi^{3}_{1}= \mathcal{P}_{\rm LLL} |\Phi_{n}|^2 \prod_{j<k}(z_j-z_k)^3.
\label{eq:parton1}
\end{equation}
Here we have suppressed the Gaussian factors for simplicity and have used $\Phi_{1}\sim \prod_{j<k}(z_j-z_k)$. All these states occur at a filling factor $\nu=1/3$ and have a shift $\mathcal{S}=3$, which is the same as the shift of the Laughlin state at $\nu=1/3$.  One might at first think that the factor $\Phi_{n}\Phi_{\bar{n}}$ does not change the topological structure of the state. However, this factor represents condensation of clusters of $n$ bosons~\cite{Barkeshli13}, and thus endows the 1/3 state with $\mathbb{Z}_{n}$ order, making it (for $n>1$) topologically distinct from the Laughlin state. A very nice physical interpretation of $\Psi^{n\bar{n}111}_{1/3}$ follows from the composite boson theory of Zhang, Hansson and Kivelson~\cite{Zhang89}, which interprets the Laughlin wave function as a Bose condensate of composite bosons. $\Psi^{n\bar{n}111}_{1/3}$ can similarly be interpreted as a $\mathbb{Z}_n$ superconductivity of composite bosons (see schematic shown in Fig.~\ref{fig:schematic}), because the factor $\Phi^{3}_{1}\sim \prod_{j<k}(z_j-z_k)^3$ binds three vortices to each electron to convert it into a composite boson, and composite bosons then condense into the $\mathbb{Z}_n$ superconductor described by $|\Phi_{n}|^2$. (For the Laughlin state, the wave function of the condensed bosons is simply 1~\cite{Zhang89}.) $\mathbb{Z}_{n}$ order implies that as we add composite bosons one by one, adding $n$ composite bosons is equivalent to adding no composite bosons since these become part of the condensate. An indication of the topological distinction of the different $\mathbb{Z}_n$ states can be seen from the fact that the $n\bar{n}111$ state can be constructed, on a compact geometry such as the sphere, only when $N$ is divisible by $n$. The $1\bar{1}111$ state, which can be constructed for arbitrary $N$, is seen below as equivalent to the Laughlin state.

The second motivation is experimental. As noted above, the 1/3 state in the LLL was the first to be seen and explained. One might expect analogous physics for FQHE at $\nu=1/3$ in the SLL of a GaAs system, which appears at $\nu=7/3$~\cite{Willett87} (with the lowest filled Landau level of up and down spins contributing 2 to the filling factor). However, Laughlin's wave function for the 7/3 FQHE is not fully convincing for the following reasons: (i) The overlap of the exact SLL Coulomb ground state with the Laughlin state is $<0.6$ for systems up to $N=15$ electrons~\cite{Ambrumenil88,Balram13b,Kusmierz18,Balram20}. This should be contrasted with the overlap of the exact LLL Coulomb ground state with the Laughlin state which is $>0.98$ for systems up to $N=15$~\cite{Ambrumenil88,Kusmierz18,Balram20}.
(ii) The quasihole and quasiparticle excitations at 7/3 do not appear at the quantum numbers expected from the Laughlin description~\cite{Balram13b}. (iii) There is no indication of a low-energy magnetoroton mode~\cite{Girvin85,Girvin86} in the exact spectra at 7/3 akin to the one that can be clearly identified at 1/3. On the contrary, it seems that there may be evidence of multiple branches of neutral excitations in both exact diagonalization~\cite{Balram13b,Jolicoeur17} and resonant inelastic light scattering experiments~\cite{Wurstbauer15}. Owing to these facts, the precise nature of the $7/3$ state has been a matter of debate in the literature~\cite{Balram13b,Johri14,Zaletel15,Peterson15,Jeong16}. 

These observations raise the question: Could the $n\bar{n}111$ states be relevant for the 7/3 FQHE?  That has prompted us to carry out a quantitative investigation of the issue. Our primary findings are as follows:

{\it Quantitative plausibility:}  The $2\bar{2}111$ and $3\bar{3}111$ states have a higher overlap than the Laughlin state with the SLL Coulomb ground state for systems for which the overlaps can be evaluated. Furthermore, the $2\bar{2}111$ and $3\bar{3}111$ states have slightly lower energy than the Laughlin state in the thermodynamic limit at $\nu=7/3$. (In contrast, in the LLL the Laughlin state is clearly superior to the $2\bar{2}111$ and $3\bar{3}111$ states.)  We therefore conclude that the $2\bar{2}111$ and $3\bar{3}111$ are at least as plausible for the 7/3 FQHE as the Laughlin state. Since the energy differences are too small to be decisive, it is important to identify experiments that that can look for potential signatures of $\mathbb{Z}_n$ order at $\nu=7/3$.

{\it Splitting the Laughlin quasihole:}  Perhaps the most remarkable aspect of the $n\bar{n}111$ state is that it hosts quasiparticles with charge of magnitude $e/(3n)$. There are two closely related ways to see this. From the wave function perspective, consider the Laughlin-quasihole located at $\eta$, obtained by multiplication of the ground state by the factor $\prod_j(z_j-\eta)$. This has the usual charge $e/3$. However, when viewed in conjunction with $\Phi_n$, the state $\prod_j(z_j-\eta)\Phi_n$ is not a single hole but actually a collection of $n$ holes in the factor $\Phi_n$. The factor $\prod_j(z_j-\eta)$ thus produces $n$ quasiholes, which can be moved away from one another to produce {\it elementary} quasiholes of charge $e/(3n)$ each. From the field theory perspective, the $\mathbb{Z}_{n}$ superconductivity in the $n\bar{n}$ factor results in vortices with flux $h/(ne)$, which carries charge $e/(3n)$ (because a unit flux $h/e$ has charge $e/3$ associated with it). We construct below explicit wave functions for the elementary quasiparticles and confirm their charge. The existence of charge $e/(3n)$ excitations implies that the addition of a single flux quantum to the $n\bar{n}111$ state produces $n$ quasiholes. This has implications for the structure of the low-energy spectrum slightly away from 7/3, which may explain the unusual nature of the exact Coulomb spectra slightly away from 7/3.

{\it Multiple neutral excitations:}  Because it is possible to create quasiparticles and quasiholes in either the same or different factors $\Phi_n$ and $\Phi_n^*$, the $n\bar{n}111$ state is predicted to support two neutral excitation branches which are topologically distinct. The trivial exciton (with quasiparticle and quasihole in the same factor) in the spherical geometry occurs at the same quantum numbers (same $N$ and $2Q$) as the ground state, whereas the topologically non-trivial exciton (with quasiparticle and quasihole in different factors) occurs at quantum numbers which are different from those of the ground state. The topologically non-trivial neutral excitation corresponds to the composite boson, which in this state is a topologically non-trivial, electrically neutral bosonic excitation. Analogous physics has been noted for the Pfaffian state~\cite{Moller11,Sreejith11}. 

{\it Abelian Braid statistics:}  In spite of the anomalous charge, the excitations of the $n\bar{n}111$ state carry Abelian braid statistics. That is evident from the fact that the partons corresponding to $\bar{n}$ and $n$ have opposite charges and are thus distinguishable.  

{\it Field theory:} The field theory for the $n\bar{n}111$ state can be derived from the parton construction, which leads to a bulk effective Abelian Chern-Simons theory characterized by a $3 \times 3$ $K$-matrix [see Eqs.~(\ref{eq:Kmatrix_n_barn_111}) and (\ref{eq:Kmatrix2_n_barn_111})]. This field theory reveals a $\mathbb{Z}_n$ topological order in the $n\bar{n}111$ state, in the sense that the theory can be thought of as gauging the $\mathbb{Z}_n$ subgroup of the $U(1)$ global symmetry of the $1/3$ Laughlin state (see Ref.~\onlinecite{Barkeshli14a} for a general theory of gauging in topologically ordered systems). In particular, this predicts a ground state degeneracy of $3n^2$ on the torus. The minimally charged quasiparticle, with charge $-e/(3n)$, carries a fractional exchange statistics of $\theta = \pi/(3n^2) - \pi/n + \pi$ (defined modulo $2\pi$), and its fusion rules form a $\mathbb{Z}_{3n^2}$ group (see Section \ref{sec:field_theory} and Appendix~\ref{app:effective_field_theory_parton_nbarn111} for a simple counting of the number of topologically distinct quasiparticle sectors). The topologically non-trivial neutral excitation corresponds to the $\mathbb{Z}_n$ charge (equivalently, this is the composite boson). 

{\it Edge structure:}  Naively one would expect the $n\bar{n}111$ state to possess counter-propagating neutral edge modes. That is a general property of states consisting of factors of the type $\Phi_n^*$ that correspond to a negative magnetic field (because of complex conjugation); edge excitations in such a factor move in the backward direction, which combine with the other edge modes to yield counter-propagating neutral edge modes. However, unlike the backward moving modes in the $n/(2n-1)$ Jain states, the counter-propagating modes of the $n\bar{n}111$ state are not topologically robust and backscattering terms can gap these modes out (provided that the strength of the backscattering terms is sufficiently large). The only robust edge mode of the $n\bar{n}111$ state is the same as the forward propagating mode of the Laughlin $\nu=1/3$ state. In particular, gapping out the backward propagating modes also implies that it costs a finite energy at the edge to create the quasiparticles with charge $-e/(3n)$, and only quasiparticles of charge $-e/3$ (and their integer multiples) can be created at the edge at arbitrarily low energies. This is intuitively understandable because the charge $-e/(3n)$ quasiparticles are actually the vortices of the $\mathbb{Z}_{n}$ superconductor, and vortices of superconductors remain gapped at the boundary.

While the $2\bar{2}111$ and $3\bar{3}111$ states are variationally slightly better than the Laughlin state in the SLL, the energies of all of these states are so close that only experiments can conclusively establish which one occurs in nature, especially as our calculations do not include the effects of LL mixing, finite quantum well width and disorder. (Recall that changing the quantum well width or LL mixing effectively modifies the interaction.) The above listed properties can in principle allow experiments to identify the nature of the state. It is possible that more than one of these states may be realized as a function of  parameters. 

For convenience, we list in Table~\ref{tab:properties_states_1_3} the various predicted properties of the $\mathbb{Z}_n$ states, derived in detail later. The Laughlin state corresponds to $\mathbb{Z}_1$. This table also includes predictions from another candidate state at $\nu=1/3$, namely the particle-hole conjugate of the four-cluster Read-Rezayi (aRR$4$) state~\cite{Read99}.

\begin{table}[t]
\begin{center}
\begin{tabular} { | c | c| c | c | c |  c |  c |}
\hline
State &  $\mathcal{S}$ & $c_{-}$ & $\mathcal{D}$ & excitations &$\mathcal{Q}_{e}$ (bulk) & $\mathcal{Q}_{e}$ (edge)\\
\hline
$n\bar{n}111$ ($\mathbb{Z}_n$) & $3$ &  $1$  & $3n^{2}$  & Abelian 		 & $-e/(3n)$ & $-e/3$  \\ 
\hline
Laughlin & $3$ &  $1$  & $3$  & Abelian 		 & $-e/3$ & $-e/3$  \\ \hline
aRR$4$               						&$-3$ &  $-1$ &    $15 $    & non-Abelian &  $-e/6$    & $-e/6$  \\ \hline
\end{tabular}
\end{center}
\caption{\label{tab:properties_states_1_3} This table gives various topological parameters for the $n\bar{n}111$ parton states at $\nu=1/3$, which carry $\mathbb{Z}_{n}$ topological order. $\mathcal{S}$ is the shift on the sphere, $c_{-}$ is the chiral central charge, $\mathcal{D}$  is degeneracy on the torus, and $\mathcal{Q}_{e}$ (bulk) and $\mathcal{Q}_{e}$ (edge) are the charges of the elementary (with smallest magnitude charge) quasiparticle in the bulk and at the edge, respectively. $\mathcal{Q}_{e}$ (edge) refers to the minimal charge that can be created at arbitrarily low energies on the edge. Nature of the excitations (Abelian or non-Abelian) is also indicated. The shift determines the Hall viscosity of a fractional quantum Hall state as $\eta_{H}=\hbar\mathcal{S}/(24\pi\ell^{2})$, and the chiral central charge is related to the thermal Hall conductance as $\kappa_{xy}=c_{-}~[\pi^2 k_{\rm B}^2 /(3h)]T$, where $T$ is the temperature. The values of these topological quantities are also given for the Laughlin state and for the particle-hole conjugate of the four-cluster Read-Rezayi (aRR$4$) state. The $1\bar{1}111$ parton state is topologically equivalent to the Laughlin state. }
\end{table}

FQHE states with $\mathbb{Z}_{n}$ order can be constructed at other fractions in a similar fashion, i.e. by multiplying a FQHE wave function, such as the Jain wave function $m11$, by the factor $n\bar{n}$ (with $m\neq n$ and $m\neq \bar{n}$), which leaves the filling factor unchanged. We have not explored the feasibility of such states in the present study.  
 
The remainder of the paper is organized as follows. In Sec.~\ref{sec:quantitative} we present numerical results to demonstrate that the states defined in Eq.~(\ref{eq:parton1}) are viable for the actual Coulomb ground state in the SLL. Sec.~\ref{sec:charge} shows that the charge of the elementary excitation in the $n\bar{n}111$ state has magnitude $e/(3n)$. A wave function is constructed for this excitation and it is shown that it has lower energy in the SLL than the standard $e/3$ charged quasihole. In Sec.~\ref{sec:field_theory} we describe certain properties of the  $n\bar{n}111$ state which are obtained from the effective field theory of its edge. We conclude the paper in Sec.~\ref{sec:discussion} with a discussion on the experimental implications of our results and an outlook for the future.

\section{Quantitative studies}
\label{sec:quantitative}
Throughout this work we shall consider fully spin polarized electrons unless otherwise stated. Furthermore, we make the simplifying assumption of considering an ideal system with zero width, no disorder and no LL mixing. The problem of interacting electrons confined to the SLL is formally equivalent to the problem of electrons residing in the LLL but interacting via an effective interaction that has the same Haldane pseudopotentials as the Coulomb interaction in the SLL~\cite{Haldane83}. This allows us to work directly with wave functions in the LLL. 

All our numerical calculations are carried out on the Haldane sphere~\cite{Haldane83} where $N$ electrons move on a spherical surface in the presence of a radial magnetic field $B$ generated by a monopole of strength $2Q$ located at the center of the sphere. In this geometry, the total orbital angular momentum $L$ and its $z$-component $L_{z}$ are good quantum numbers. 

The wave functions stated in Eq.~(\ref{eq:parton1}) can be conveniently projected to the LLL as:
\begin{eqnarray}
\label{eq:parton2}
\Psi^{n\bar{n}111}_{1/3} &\sim & {[\mathcal{P}_{\rm LLL} \Phi_{n}\Phi_1^2] [\mathcal{P}_{\rm LLL}[\Phi_{n}]^{*}\Phi_1^2]  \over \Phi_1} \nonumber \\
&=& {\Psi^{\rm CF}_{n/(2n+1)}\Psi^{\rm CF}_{n/(2n-1)}\over \Phi_1},
\end{eqnarray}
where $\Psi^{\rm CF}_{n/(2n\pm 1)}$ is the wave function of the Jain-CF state. Eq.~(\ref{eq:parton2}) defines the LLL projection in  Eq.~(\ref{eq:parton1}) in a certain way. LLL projection can be carried out in other ways, but we expect that such details do not affect the topological properties of the underlying state~\cite{Balram16b}. (Indeed, LLL projection is not considered at all in the field theoretical description.) We will use Eq.~(\ref{eq:parton2}) in what follows below, because it allows accessibility to large systems. In this work, we shall focus on the $n=2$ and $n=3$ cases of the above wave function. We note that in the limit $n\rightarrow \infty$, the constituent wave functions on the right hand side of Eq.~(\ref{eq:parton2}) describe the CF Fermi sea. Thus the $n\bar{n}111$ state becomes compressible in the limit $n\rightarrow \infty$.

In Tables~\ref{overlaps1} and \ref{overlaps2} we show the overlap of the $2\bar{2}111$ and $3\bar{3}111$ states [see Eq.~(\ref{eq:parton2})] with both the exact SLL Coulomb ground state and the Laughlin state for small system sizes. For these small system sizes, the constituent Jain CF states of Eq.~(\ref{eq:parton2}) were obtained using a brute-force direct projection to the LLL. The tables also contain overlap of the Laughlin wave function with the exact SLL Coulomb ground state. The $2\bar{2}111$ and $3\bar{3}111$  states have a much higher overlap with the exact $7/3$ Coulomb state than the Laughlin state for all the systems we have considered. 

\begin{figure}[t]
\begin{center}
\includegraphics[width=0.43\textwidth,height=0.29\textwidth]{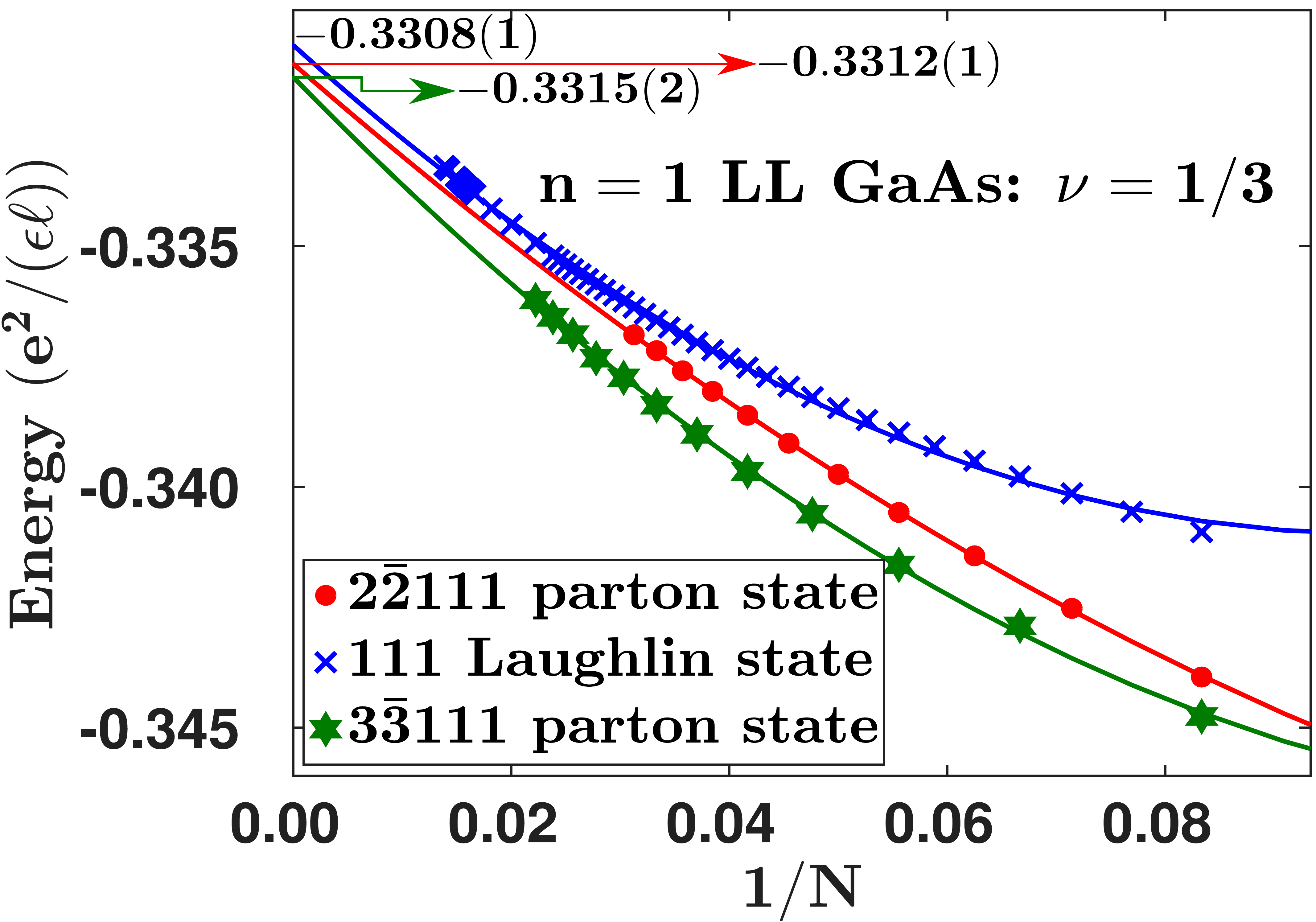} 
\caption{(color online) Thermodynamic extrapolation of the per-particle energies for the Laughlin (blue crosses), the $2\bar{2}111$ (red dots) and the $3\bar{3}111$ (green hexagrams) states at $\nu=1/3$ in the $n=1$ LL of GaAs. The energies include the electron-background and background-background contributions, and are quoted in units of $e^2/(\epsilon\ell)$. The thick lines are quadratic fits in $1/N$.}
\label{fig:extrapolations_energies_7_3}
\end{center}
\end{figure}

Overlaps for larger systems require significantly greater computer resources. However, using methods given in the literature~\cite{Jain97,Jain97b,Davenport12}, the Jain-CF states on the right hand side of Eq.~(\ref{eq:parton2}) can be constructed for very large systems. Fig.~\ref{fig:extrapolations_energies_7_3} shows the Coulomb energies of various candidate states at $\nu=7/3$ for finite systems. All the numbers quoted in Fig.~\ref{fig:extrapolations_energies_7_3} include the electron-background and background-background contributions and have been density corrected~\cite{Morf86}. We find that the $2\bar{2}111$ and $3\bar{3}111$ state have lower energies than the Laughlin state at $\nu=7/3$ in the thermodynamic limit. We note that Ref.~\cite{Jeong16} quotes the energy of the Laughlin state at $7/3$ to be $-0.325(0)~e^2/(\epsilon\ell)$ ($\ell=\sqrt{\hbar/(eB)}$ is the magnetic length), which is higher than the value shown in Fig.~\ref{fig:extrapolations_energies_7_3}. This discrepancy stems from slight differences in the form of the effective interaction used to simulate the physics of the SLL. The form of the effective interaction we use is more accurate than the form used in Ref.~\cite{Jeong16}. 

\begin{figure}[ht]
\begin{center}
\includegraphics[width=0.45\textwidth,height=0.3\textwidth]{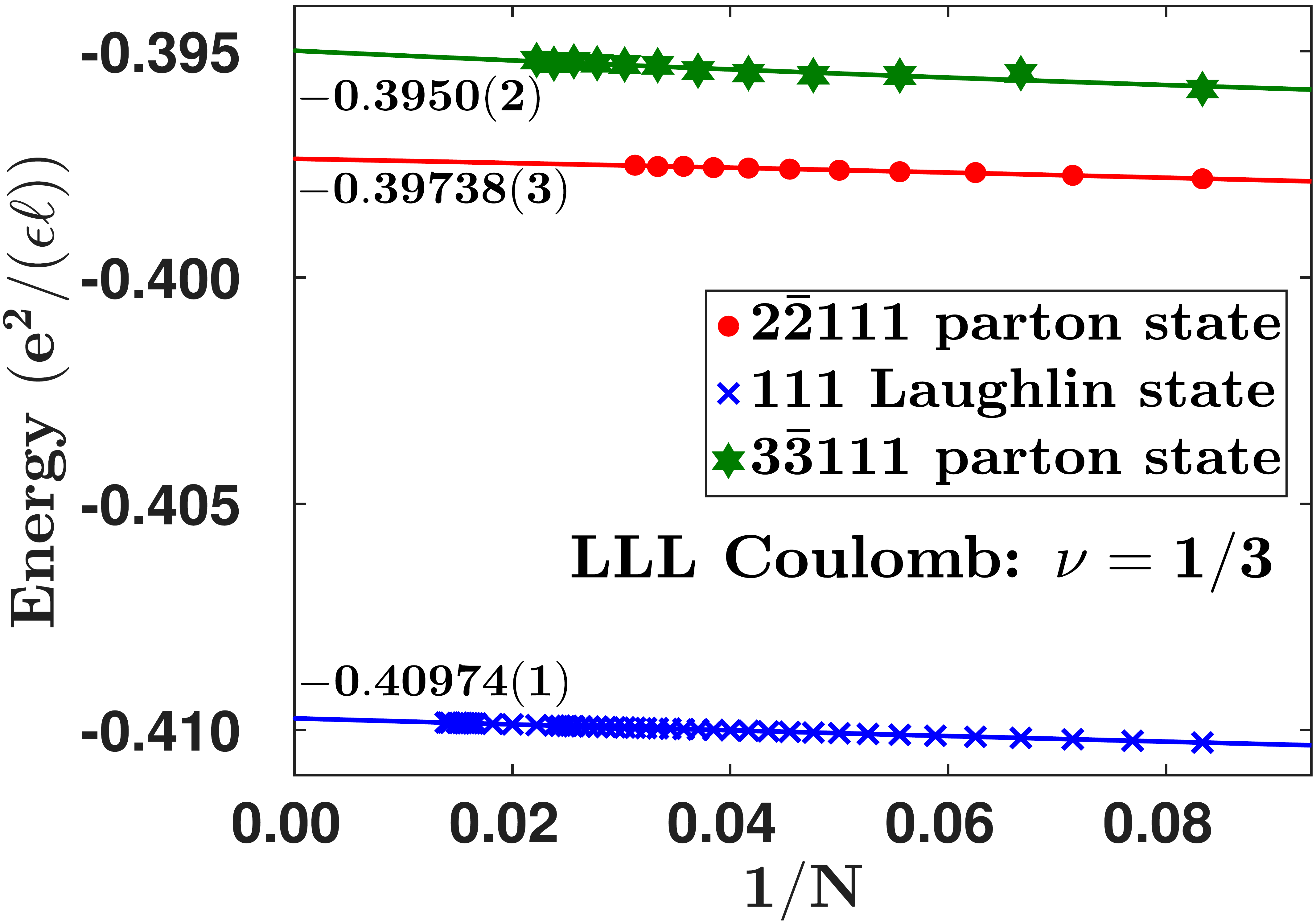} 
\caption{(color online) Same as in Fig.~\ref{fig:extrapolations_energies_7_3} but for $\nu=1/3$ in the $n=0$ LL.}
\label{fig:extrapolations_energies_1_3}
\end{center}
\end{figure}

We have also performed the comparison for the 1/3 state in the LLL, presented in Fig.~\ref{fig:extrapolations_energies_1_3}. As anticipated, the Laughlin state has lower energy than the $2\bar{2}111$ and $3\bar{3}111$ states. 

These comparisons demonstrate that the $2\bar{2}111$ and $3\bar{3}111$ states are strong candidates for the 7/3 FQHE. It is therefore important to ask in what measurable way they differ from the Laughlin state, and also, how their new topological orders manifests itself.

\begin{table}
\centering
\begin{tabular}{|c|c|c|c|c|}
\hline
$N$ & $2Q$ & $|\langle \Psi^{1{\rm LL}}_{1/3}|\Psi^{\rm Laughlin}_{1/3}\rangle|$ &  $|\langle \Psi^{2\bar{2}111}_{1/3}|\Psi^{\rm Laughlin}_{1/3}\rangle|$ & $|\langle \Psi^{1{\rm LL}}_{1/3}|\Psi^{2\bar{2}111}_{1/3} \rangle|$ 
\\ \hline
4  & 9   		&0.4765    & 0.7467 &	0.9406	 \\ \hline
6  & 15  		&0.5285    & 0.7144 &	0.9507	 \\ \hline
8  & 21 	 	&0.5720    & 0.6246 &	0.9755   \\ \hline
10 & 27 	 	&0.5400    & 0.5434 & 	0.8682	 \\ \hline
\end{tabular}
\caption{\label{overlaps_n_1_LL_2_3_2_5_over_1_parton} Overlaps of the ground state at the Laughlin flux in the $n=1$ Landau level (obtained by exact diagonalization), $\Psi^{1{\rm LL}}_{1/3}$, with the Laughlin, $\Psi^{\rm Laughlin}_{1/3}$, and $\Psi^{2\bar{2}111}_{1/3}$ states. The numbers for $|\langle \Psi^{1{\rm LL}}_{1/3}|\Psi^{\rm Laughlin}_{1/3} \rangle|$ were previously given in Ref.~\cite{Ambrumenil88} and have been reproduced here for comparison.}
\label{overlaps1}
\end{table}

\begin{table}
\centering
\begin{tabular}{|c|c|c|c|c|}
\hline
$N$ & $2Q$ & $|\langle \Psi^{1{\rm LL}}_{1/3}|\Psi^{\rm Laughlin}_{1/3} \rangle|$ &  $|\langle \Psi^{3\bar{3}111}_{1/3}|\Psi^{\rm Laughlin}_{1/3} \rangle|$ & $|\langle \Psi^{1{\rm LL}}_{1/3}|\Psi^{3\bar{3}111}_{1/3} \rangle|$ 
\\ \hline
9  & 24   		& 0.4794    & 0.4604 &	0.9256	 \\ \hline
\end{tabular}
\caption{\label{overlaps_n_1_LL_3_5_3_7_over_1_parton} Overlaps of the second Landau level Coulomb ground state, $\Psi^{1{\rm LL}}_{1/3}$, with the Laughlin and $\Psi^{3\bar{3}111}_{1/3}$ states.}
\label{overlaps2}
\end{table}

\section{Charge and entanglement spectrum}
\label{sec:charge}
Perhaps the most dramatic feature of the new states is that their elementary quasiparticle/quasihole has a charge of magnitude $e/(3n)$.  We note that this is the charge of the corresponding parton $e_\alpha=-\nu e/n_\alpha=-e/(3n)$.  Another way to obtain the charge is to consider insertion of a single flux quantum to the $n\bar{n}111$ state at location $\eta$. Following Laughlin, this results in a state described by the wave function $\prod_{j}(z_j-\eta) \Psi^{n\bar{n}111}$, and has a charge $e/3$ associated with it. However, the product $\prod_{j}(z_j-\eta) \Phi_n$ actually represents $n$ elementary holes in the state at $\nu=n$, one in each Landau level, which can be separated from one another. The state $\prod_{j}(z_j-\eta) \Psi^{n\bar{n}111}$ thus represents $n$ elementary quasiholes, each with charge $e/(3n)$. We next construct an explicit wave function for the charge $e/(3n)$ quasihole by taking the example of the $2\bar{2}111$ state.

There are two inequivalent ways of creating quasiholes carrying charge $e/6$ in the $2\bar{2}111$ state:
In the first one we create two holes in $\Phi_{2}$. The wave function of this state is given by:
 \begin{equation}
  \Psi^{\rm qh}_{1/3} = \mathcal{P}_{\rm LLL}\Phi^{2\text{-}{\rm h}}_{2} [\Phi_{2}]^{*}\Phi^{3}_{1}  \sim \frac{\Psi^{2\text{-}{\rm qh}}_{2/5}\Psi_{2/3}}{\Phi_{1}},
  \label{eq:wf_2quasiholes_2_5}
 \end{equation}
 where $\Phi^{2\text{-}{\rm h}}_{2}$ and $\Psi^{2\text{-}{\rm qh}}_{2/5}$ are the $\nu=2$ IQH and $\nu=2/5$ CF states with two holes or two quasiholes, respectively. 
 Alternatively, we create two particles in $[\Phi_{2}]^{*}$. The wave function of this state is given by:
 \begin{equation}
  \Psi^{\rm qh}_{1/3} = \mathcal{P}_{\rm LLL}\Phi_{2} [\Phi^{2\text{-}{\rm p}}_{2}]^{*}\Phi^{3}_{1}  \sim \frac{\Psi_{2/5}\Psi^{2\text{-}{\rm qh}}_{2/3}}{\Phi_{1}},
  \label{eq:wf_2quasiholes_2_3}
 \end{equation}
 where $\Phi^{2\text{-}{\rm p}}_{2}$ is the $\nu=2$ IQH state with two particles. (Here we have used the fact that reverse-vortex attachment converts quasiparticles into quasiholes and vice versa.)

\begin{figure}[h]
\begin{center}
\includegraphics[width=1.0\columnwidth]{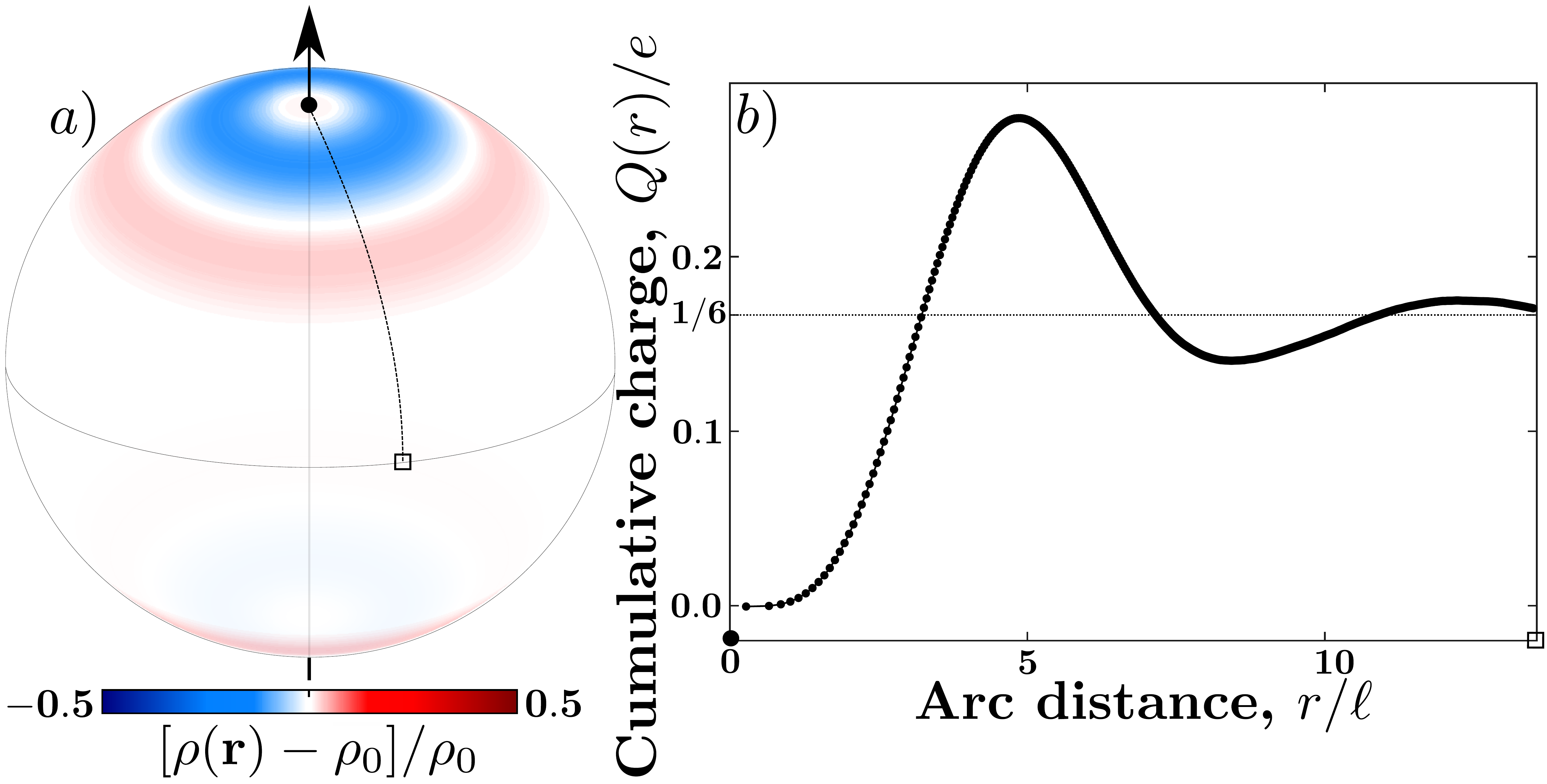}  
\caption{a) Density profile $\rho({\bf r})$ of a state with two far-separated quasiholes at $\nu=1/3$ modeled by the wave function given in Eq.~(\ref{eq:wf_2quasiholes_2_5}) for $N=50$ electrons on the sphere. The two quasiholes are located at the opposite poles of the sphere. The quantity shown is $[\rho({\bf r})-\rho_{0}]/\rho_{0}$, where $\rho_{0}$ is the density of the uniform $2\bar{2}111$ state at $\nu=1/3$. b) The cumulative charge $Q(r)$ (defined in the text) as a function of the distance $r$ (in units of the magnetic length $\ell$) measured along the arc from the north pole to the equator. The cumulative charge approaches the value $0.1\bar{6}e$ ($-e$ is the electron charge) near the equator.}
\label{fig:density_2fquasiholes_22bar111}
\end{center}
\end{figure}

A nice feature of our wave functions is that they can be constructed for very large systems. We now demonstrate that the elementary quasihole of the $2\bar{2}111$ state indeed carries charge $e/6$. In Fig.~\ref{fig:density_2fquasiholes_22bar111}a) we show the density profile $\rho({\bf r})$ of the $2\bar{2}111$ state at $\nu=1/3$ for $N=50$ electrons with two far-separated quasiholes, located at the two poles of the sphere. We model this state using the wave function given in Eq.~(\ref{eq:wf_2quasiholes_2_5}), with the LLL projection evaluated using the Jain-Kamilla method~\cite{Jain97,Jain97b,Moller05,Jain07,Davenport12,Balram15a}. Away from the poles, the density of the state relaxes to the density $\rho_{0}$ of the uniform state $2\bar{2}111$. In Fig.~\ref{fig:density_2fquasiholes_22bar111}b) we plot the cumulative charge $Q(r)=(-e)\int^{r}_{0}d^{2}{\bf r'}[\rho({\bf r'})-\rho_{0}]$, where $r$ parametrizes the latitude on the sphere in terms of the arc distance measured from the north pole. For the system of $N=50$ electrons shown in Fig.~\ref{fig:density_2fquasiholes_22bar111}, we obtain a cumulative charge of $0.17e$ near the equator which is close to the expected value of $0.1\bar{6}e$. The slight discrepancy arises because of the slight overlap between the quasiholes at the north and the south poles and should go away for larger systems. It is noted that the quasihole of the $2\bar{2}111$ state is more spread out than that of the Laughlin state.

\begin{figure}[h]
\begin{center}
\includegraphics[width=0.45\textwidth,height=0.3\textwidth]{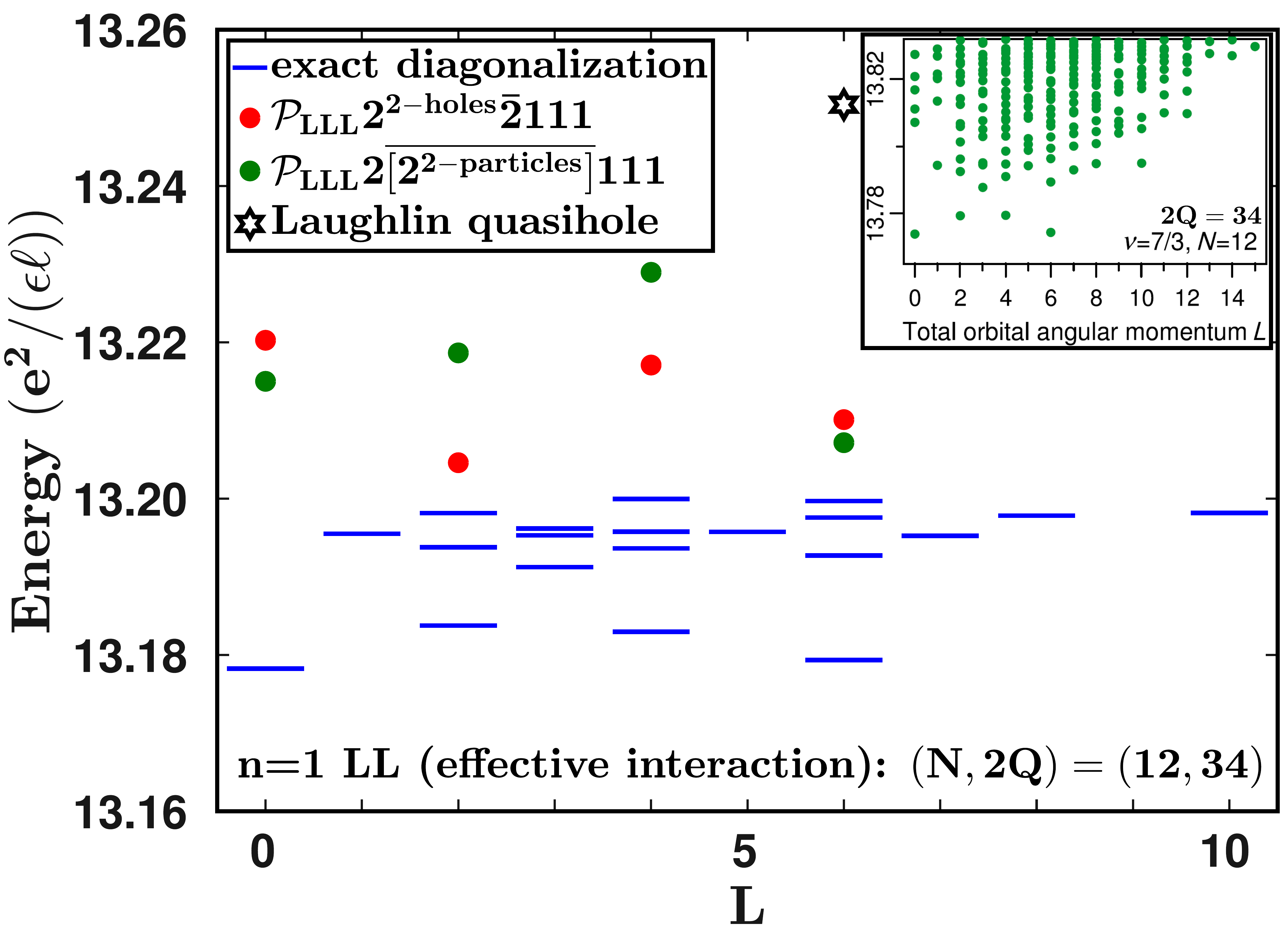}
\caption{(color online) Spectrum of the effective interaction~\cite{Shi08} which simulates the the physics of the second Landau level (LL) in the lowest LL (LLL) for $N=12$ electrons at a flux $2Q=34$ obtained by exact diagonalization in the spherical geometry [blue lines]. This flux corresponds to a single quasihole state (see text) at $7/3$ which in the $2\bar{2}111$ state can be represented by two wave functions given in Eqs.~(\ref{eq:wf_2quasiholes_2_5}) [red dots] and (\ref{eq:wf_2quasiholes_2_3}) [green dots] and in the Laughlin construction by a single quasihole [black hexagram]. The $2\bar{2}111$ state gives a decent description of the low-energy structure of the single quasihole state while the Laughlin quasihole has a much higher energy. For comparison we also show the exact second LL Coulomb spectra reproduced from Ref.~\cite{Balram13b} as an inset which demonstrates that the low-energy structure of the exact second LL Coulomb and the effective interaction spectra are similar. The energy of the Laughlin quasihole with the exact second LL Coulomb interaction for this system is 13.840 (off the chart).}
\label{fig:7_3_quasihole}
\end{center}
\end{figure}

The presence of unconventional quasiholes/quasiparticles also implies a different low energy spectrum at $7/3$ compared to that in the vicinity of the LLL 1/3 state. In Fig.~\ref{fig:7_3_quasihole} we show the exact spectrum for the SLL system at $\nu=1/3$ with an additional flux. The calculation is done for $N=12$ electrons in the spherical geometry. An effective interaction~\cite{Shi08} has been used for the calculation which simulates the physics of the SLL in the LLL. Upon the addition of a single flux quantum in the the LLL 1/3 state, one would obtain a single state at $L=6$. In contrast, insertion of a single flux quantum at $\nu=7/3$ results in a low energy sector consisting of multiplets at $L=0, 2, 4, 6$. Remarkably, these are precisely the quantum numbers predicted for two holes in $\Phi_2$ or two particles in $\Phi^*_2$. The $2\bar{2}111$ description thus captures a remarkable feature of the exact spectra, namely that its low energy sector consists of multiplets at $L=0, 2, 4, 6$.  We construct wave functions for these multiplets and obtain their energies, also shown in Fig.~\ref{fig:7_3_quasihole}. The parton theory is seen to give reasonable variational energies. Note we have two sets of energies, depending on whether we consider holes in $\Phi_2$ or particles in $\Phi^*_2$. We expect that the energies can be further improved by doing CF Diagonalization (CFD)~\cite{Mandal02} in the subspace of these two states. In contrast, the Laughlin quasihole is a single state that occurs at $L=N/2$ and has a significantly higher energy.

One might wonder about the braid statistics of the quasiparticles of the $n\bar{n}111$ states. Wen demonstrated~\cite{Wen91} that the quasiparticle excitations of the Jain parton states of the form $nn n_{3}n_{4}\cdots$, which contain repeated integers $n>1$, carry non-Abelian braid statistics. These result from the fact that the partons from the two factors of $\Phi_n$ are identical, and thus the physical state must be invariant under an $SU(2)$ rotation within the parton space. (Repeated $1$'s do not give non-Abelian quasiparticles.) For the present case, the partons corresponding to $\Phi_n$ and $\Phi_n^*$ have different charges. As a result, the excitations of the $n\bar{n}111$ states are Abelian.

\begin{figure}[ht]
\begin{center}
\includegraphics[width=0.43\textwidth]{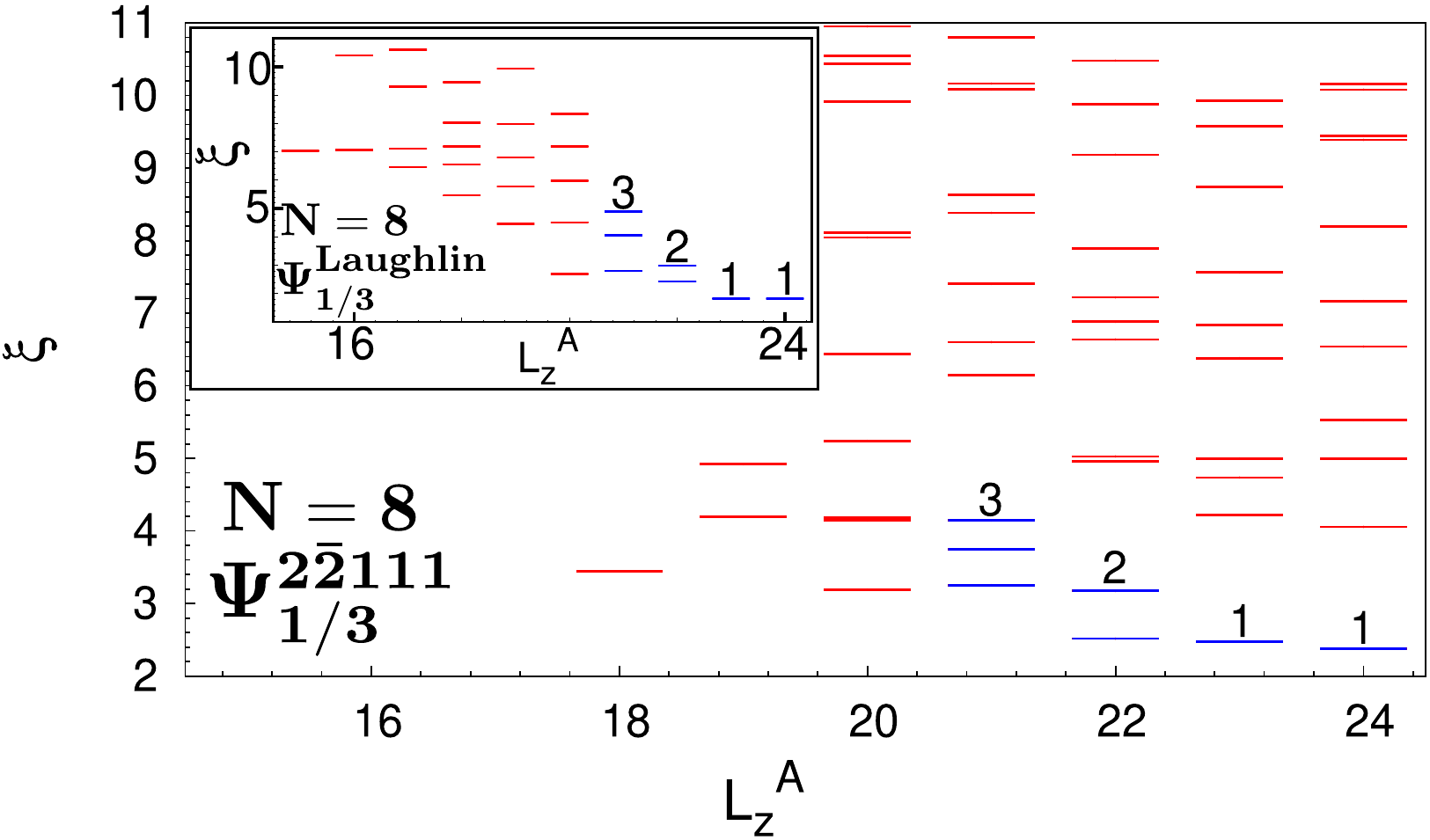}\\
\includegraphics[width=0.43\textwidth]{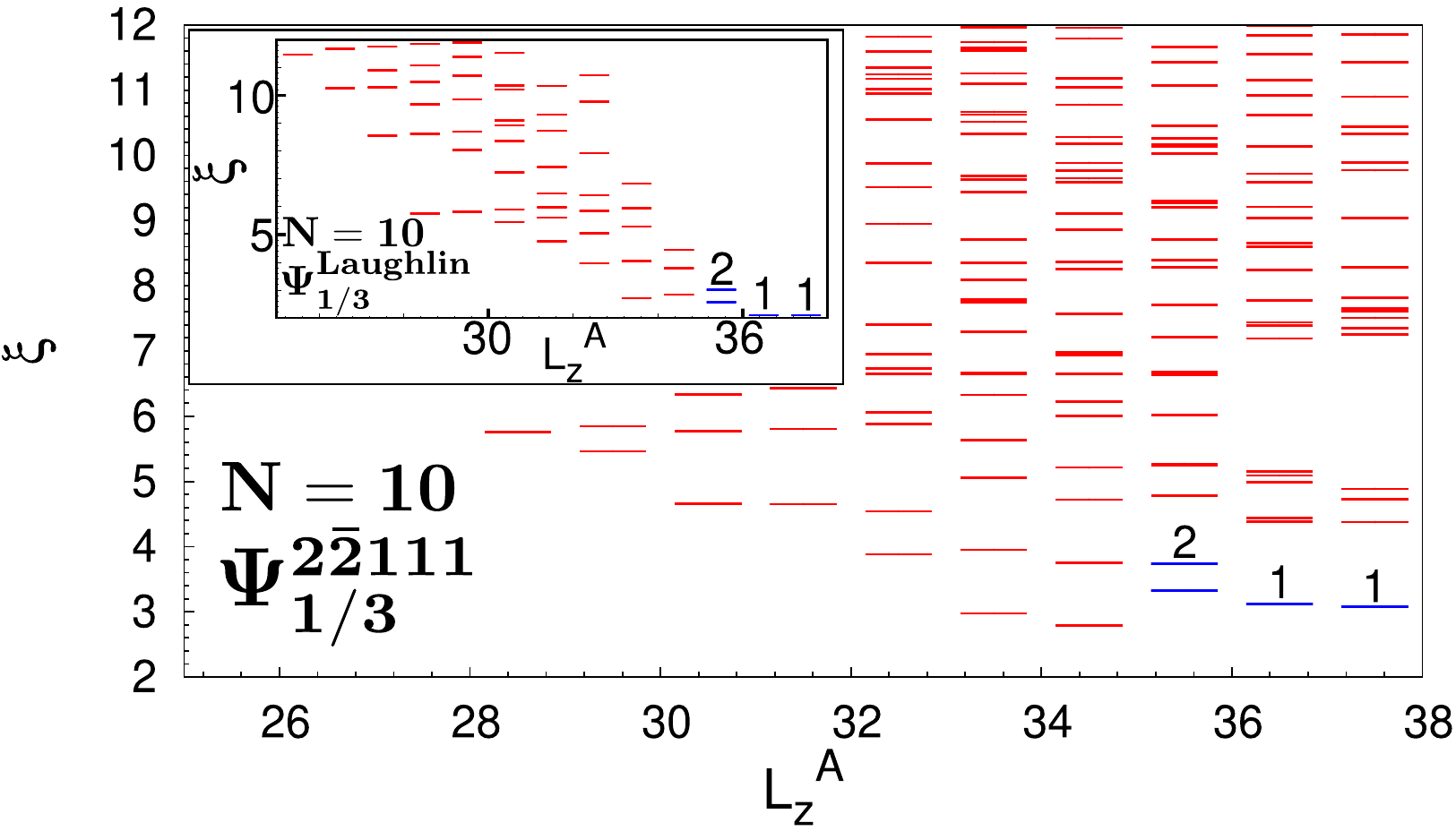}
\caption{(color online) Top (bottom) panel shows the orbital entanglement spectrum for an equal bipartition of the $\Psi^{2\bar{2}111}_{1/3}$ state for $N=8$ ($N=10$) electrons at a flux $2Q=21$ ($2Q=27$) on the sphere. The entanglement levels are labeled by the $z$-component of the total orbital angular momentum of the $A$ subsystem, $L_{z}^{A}$. For comparison we also show the corresponding spectra for the Laughlin state (insets). The counting of low-lying levels for $N=8$ (from $L_{z}^{A}=24$ going from right to left) goes as $1,1,2,3\cdots$, and is identical for the two states. The counting for $N=10$ is not quite definitive.}
\label{fig:entanglement_spectra_entspec_Laughlin_2bar2111}
\end{center}
\end{figure}

Next, we compare the entanglement spectrum of the $2\bar{2}111$ state with the Laughlin state. The entanglement spectrum has served as a useful tool to characterize many FQH states since it captures the edge structure of the state which, in turn, carries a fingerprint of the topological order of the underlying state~\cite{Li08}. In Sec. \ref{edgeSec} we show that the $n \bar{n} 111$ states are described by the same edge theory at low energies, and therefore we expect the same low-lying entanglement spectrum for all $n$. In Fig.~\ref{fig:entanglement_spectra_entspec_Laughlin_2bar2111} we show the orbital entanglement spectrum~\cite{Haque07} of the $2\bar{2}111$ state for $N=8$ and $N=10$ particles. The counting of the low-lying entanglement levels for the $2\bar{2}111$ state matches with the Laughlin state for $N=8$, but for $N=10$ the counting is less clear. It is therefore not possible to ascertain the topological order of the state from the entanglement spectra of the systems that are accessible to us. 

\section{Effective field theory description of $n\bar{n}111$ states}
\label{sec:field_theory}

The difference in the fractional charges for the elementary excitations indicates that the $n\bar{n}111$ states with different $n$ are topologically distinct. In this section we consider the effective field theory that describes the topological properties of these states. A comprehensive discussion is left for Appendices ~\ref{app:effective_field_theory_parton_nbarn} and \ref{app:effective_field_theory_parton_nbarn111}, while below we summarize the salient results. We will see how the $n\bar{n}111$ state has a hidden $\mathbb{Z}_n$ topological order, which can be understood as starting with the Laughlin-$1/3$ state and gauging a $\mathbb{Z}_n$ subgroup of the global $U(1)$ charge conservation symmetry. 

First, we note that the state $n\bar{n}111$ can be understood through a parton construction
\begin{align}
\wp = b f,
\end{align}
where $\wp$ is the electron operator, $b$ is a boson, and $f$ is a fermion. The boson $b$ forms the state $n \bar{n}$, while the fermion $f$ forms the state $111$, which is simply the usual Laughlin-$1/3$ state. In this construction, $b$ can be identified with the usual composite boson of Zhang, Hansson, Kivelson~\cite{Zhang89}. The above parton construction has a $U(1)$ gauge symmetry associated with the transformations $b \rightarrow e^{i\Theta} b$, $f \rightarrow e^{-i\Theta} f$, which keeps invariant all physical operators. This gauge symmetry can be enforced in the effective field theory by introducing an emergent $U(1)$ gauge field $\beta$ that couples to $b$ and $f$ with charge $1$ and $-1$, respectively.

The electron $\wp$ has unit charge under the background physical electromagnetic field $A$. Without loss of generality, we assign charge $0$ to $b$ and $1$ to $f$ under $A$. Therefore, the effective field theory for the system can be written as
\begin{align}
\mathcal{L}_{n\bar{n}111} = \mathcal{L}_b(\beta) + \mathcal{L}_f(\beta, A) , 
\label{eq_Lagrangian_boson_plus_fermion}
\end{align}
where $\mathcal{L}_b$ and $\mathcal{L}_f$ are the Lagrangians for the $b$ and $f$ sectors separately. 
Since $f$ forms a $1/3$ Laughlin FQH state and carries unit charge with respect to the external electromagnetic field $A$, we therefore have (for convenience we set $\hbar=e=c=1$)
\begin{align}
\mathcal{L}_f = - \frac{3}{4\pi}  \tilde{a} d \tilde{a}  - \frac{1}{2\pi} (A  - \beta) d \tilde{a}, 
\label{eq:Lagrangian_density_Laughlin_1_3}
\end{align}
where we have used the notation $A d A \equiv \epsilon^{\mu\nu\lambda} A_\mu \partial_\nu A_\lambda$, we have used Einstein's summation convention and $\epsilon^{\mu\nu\lambda}$ is the fully anti-symmetric Levi-Civita tensor. The current of the $f$ fermion $j_\mu^{(f)}$ is described in terms of the emergent $U(1)$ gauge field $\tilde{a}$ as $j_\mu^{(f)} = \frac{1}{2\pi} \epsilon_{\mu\nu\lambda} \partial_\nu \tilde{a}_\lambda$. 

Let us now consider the properties of the $n\bar{n}$ state, which is the state formed by the boson $b$. This state itself can be understood through a parton construction
\begin{align}
b = f_1 f_2,
\end{align}
where $f_1$ and $f_2$ are each fermions forming a $\nu = n$ and $-n$ IQH state, respectively. This parton construction has a $U(1)$ gauge symmetry associated with the transformations $f_1 \rightarrow e^{i\theta} f_1$ and $f_2 \rightarrow e^{-i\theta} f_2$, which keeps $b$ invariant. This gauge symmetry can be enforced in the effective field theory by introducing a $U(1)$ gauge field $\alpha$ that couples to $f_1$ and $f_2$ with charge $1$ and $-1$ respectively. Furthermore, $b$ carries unit charge under $\beta$; we can assign charge $1$ to $f_1$ and charge $0$ to $f_2$ under $\beta$. As shown in Ref.~\cite{Barkeshli13} and reviewed in Appendix~\ref{app:effective_field_theory_parton_nbarn}, this construction describes a many-body state where $b^n$ has condensed. That is, the state $n \bar{n}$ is a state where the global $U(1)$ number conservation of $b$ has been broken down to $\mathbb{Z}_n$. Furthermore, an effective field theory for this state can be obtained by integrating out the $f_1$ and $f_2$ fermions:
\begin{equation}
 \mathcal{L}_b = \frac{n}{2\pi} \beta d \alpha + \frac{n}{4\pi} \beta d \beta - \frac{1}{g^{2}} (\epsilon_{\mu\nu\lambda}\partial_{\nu}\alpha)^{2},
\label{eq:Lagrangian_density_parton_n_barn}
\end{equation}
where the last term is the Maxwell action for $\alpha$ with the coupling constant $1/g^2$. The current of the $b$ bosons $j_\mu^{(b)}$ is given in terms of the emergent $U(1)$ gauge field $\alpha$ as $j_\mu^{(b)} = \frac{n}{2\pi} \epsilon_{\mu\nu\lambda} \partial_\nu (\alpha_\lambda + \beta_\lambda)$. The massless photon described by fluctuations of $\alpha$ corresponds to the Goldstone mode arising from spontaneously breaking the $U(1)$ number conservation of $b$. The second term on the right hand side of Eq.~(\ref{eq:Lagrangian_density_parton_n_barn}), $ \frac{n}{4\pi} \beta d \beta $, arises from the $\beta$-charge assignments of $f_1$ and $f_2$, and indicates a non-trivial topological response term for the $\mathbb{Z}_n$ superconducting state of $b$. Since the state of $b$ possesses a gapless Goldstone mode, it is not clear that this topological response term has well-defined physical consequences for the many-body state of $b$. However, when $b$ and $f$ are combined to form the state $n \bar{n} 111$, the Goldstone mode is gapped, and this term does play an important role in dictating the fractional statistics of the quasiparticles. 

Thus, we see that the $n \bar{n} 111$ state can be thought of a state where $b$ forms a $\mathbb{Z}_n$ condensate, while $f$ forms a $1/3$-Laughlin state. 
Combining everything following Eqs.~(\ref{eq_Lagrangian_boson_plus_fermion}), (\ref{eq:Lagrangian_density_Laughlin_1_3}) and (\ref{eq:Lagrangian_density_parton_n_barn}), we see that the effective theory of the $n\bar{n}111$ state is given by
\begin{align}
  \label{nbarn111EFT}
  \mathcal{L}_{n\bar{n}111} = &- \frac{3}{4\pi} \tilde{a} d \tilde{a}  - \frac{1}{2\pi}  (A  - \beta) d \tilde{a} +\frac{n}{2\pi} \beta d \alpha + \frac{n}{4\pi} \beta d \beta.
\end{align}
Note that when $b$ and $f$ are coupled by the gauge field $\beta$, the gapless mode associated with fluctuations of $\alpha$ becomes gapped, and the Maxwell term for $\alpha$ becomes an irrelevant subleading correction, which we have thus ignored. Therefore, the $n\bar{n}111$ state describes a gapped FQH state for a finite $n$. 

We note that this theory can be directly interpreted as a theory where we have gauged the $\mathbb{Z}_n$ subgroup of the $U(1)$ global symmetry of a $1/3$ Laughlin state. A $\mathbb{Z}_n$ gauge theory can be described by a $U(1) \times U(1)$ mutual Chern-Simons theory $\mathcal{L}_{\mathbb{Z}_n} = (n/(2\pi)) \beta d \alpha$, where $\beta$ and $\alpha$ are emergent $U(1)$ gauge fields. Therefore coupling $\beta$ to a conserved current $j_\mu^{(f)}$ can be interpreted as gauging the $\mathbb{Z}_n$ subgroup of the $U(1)$ symmetry associated with the conserved current $j_\mu^{(f)}$.

Equivalently, following the manipulations of Eqs.~(\ref{eq:Lagrangian_density_parton_n_barn_AH_mechanism_appendix})-(\ref{eq:Lagrangian_density_charge_n_SC_appendix}) (with $\beta$ appropriately replacing $A$ in those equations) and integrating out the $\tilde{a}$ gauge field, we can rewrite this effective theory as
\begin{align}
  \mathcal{L}_{n\bar{n}111} = &\frac{1}{3} \frac{1}{4\pi} (A- \beta) d (A- \beta) + \frac{n}{4\pi} \beta d \beta
  \nonumber \\
 & - c |\partial \varphi -  n \beta|^2 ,
\label{eq:Lagrangian_density_parton_nbarn111_composite_bosons} 
\end{align}
where $c$ is a constant. The scalar field $\varphi$ can be interpreted as the phase of the superconducting order parameter, $b^n$.

This form brings out the connection with the Zhang-Hansson-Kivelson (ZHK) theory, defined by
  \begin{equation}
    \label{ZHK}
  \mathcal{L}_{\rm ZHK} = 
- c |\left(\partial \varphi - n \left(\beta+A\right) \right) |^2 + \frac{1}{3} \frac{1}{4\pi} \beta d \beta\;.
\end{equation}
After translation $\beta\rightarrow \beta-A$, this becomes topologically identical to Eq.~(\ref{eq:Lagrangian_density_parton_nbarn111_composite_bosons}) with $n=1$. (The additional $1/(4\pi)\beta d\beta$ term in Eq.~(\ref{eq:Lagrangian_density_parton_nbarn111_composite_bosons}) as compared with Eq.~(\ref{ZHK}) does not change the topological properties, as seen in explicit calculations below. As discussed in the Appendices~\ref{app:effective_field_theory_parton_nbarn} and \ref{app:effective_field_theory_parton_nbarn111}, in our description the vortices of the superconductor of $b$ correspond to particle or hole excitations in the parton Landau levels, and as such are fermions before being dressed by the Chern-Simons gauge fields. In the usual ZHK description, the vortices are treated as bosons before being dressed by the Chern-Simons gauge fields; this difference in description is related to the additional $1/(4\pi)\beta d\beta$ term in the $n = 1$ case). 

As described in additional detail in the Appendices~\ref{app:effective_field_theory_parton_nbarn} and \ref{app:effective_field_theory_parton_nbarn111}, we can rewrite the field theory of the $n\bar{n}111$ state, Eq.~(\ref{nbarn111EFT}), in terms of the following Lagrangian density~\cite{Wen92b,Wen95,Moore98}:
\begin{equation}
 \mathcal{L} = -\frac{1}{4\pi}K^{(f)}_{\rm IJ}\epsilon^{\mu\nu\lambda}a_{\mu}^{\rm I}\partial_{\nu} a_{\lambda}^{\rm J}
 -\frac{1}{2\pi} t_{\rm I}\epsilon^{\mu\nu\lambda}A_{\mu}\partial_{\nu} a_{\lambda}^{\rm I}.
\end{equation}
In the above equation $A_{\mu}$ is the external electromagnetic field while $a_{\mu}$'s denote the internal gauge fields. The symmetric integer-valued $K$ matrix for the $n\bar{n}111$  state is given by
\begin{equation}
K^{(f)} =  
\begin{pmatrix} 
      3 & -1 & 0 \\
      -1 & -n & -n \\
      0 & -n  & 0 \\
   \end{pmatrix},
\label{eq:Kmatrix_n_barn_111}   
\end{equation}
and the charge vector $\vec{t}=(1,0,0)^{\rm T}$. In this notation $a^1 = \tilde{a}$, $a^2 =\beta$, and $a^3 = \alpha$.  To deduce the fractional statistics of quasiparticles, we must be careful to keep track of the fact that the quasiparticles correspond to particles / holes of the parton LLs, which are fermions that are subsequently dressed by the Chern-Simons gauge fields. As discussed in Appendix \ref{app:effective_field_theory_parton_nbarn111}, one can keep track of this by using, instead of $K^{(f)}$, a slightly different $K$-matrix:
\begin{equation}
K =  
\begin{pmatrix} 
      3 & -1 & 0 \\
      -1 & -n(n-1) & -n \\
      0 & -n  & 0 \\
   \end{pmatrix},
\label{eq:Kmatrix2_n_barn_111}   
\end{equation}
with the same charge vector as before: $\vec{t}=(1,0,0)^{\rm T}$. One can also derive an equivalent $(2n-1) \times (2n-1)$ $K$-matrix description by integrating out the constraint gauge fields $\alpha$ and $\beta$ and introducing additional $U(1)$ gauge fields to describe the currents associated with each of the parton Landau levels. This is also described in detail in Appendix~\ref{app:effective_field_theory_parton_nbarn111}. 

The ground state degeneracy of the $n\bar{n}111$  state on a manifold of genus $g$ is $|{\rm Det}(K)|^{g} =(3 n^{2})^{g}$. The quasiparticles are described by integer vectors $\vec{l}$, and carry electric charge $\mathcal{Q}_{\vec{l}} = (-e)\vec{t}^T K^{-1} \vec{l}$~\cite{Wen95}. One can verify that the quasiparticles form a $\mathbb{Z}_{3n^2}$ group under fusion, generated by the quasiparticle vector $\vec{l}^T = (0,0,-1)$, which has charge $-e/(3n)$. 

One way to understand the appearance of the $-e/(3n)$ charge from the above point of view is as follows. The state associated with $b$, since it corresponds to condensation of clusters of $n$ bosons $b$, supports a $\mathbb{Z}_n$ vortex. This $\mathbb{Z}_n$ vortex carries $2\pi/n$ flux of the $\beta$ gauge field. The flux of the $\beta$ gauge field is also felt by the $f$ fermions, which are in a $1/3$-Laughlin state. The addition of $2\pi/n$ flux into the $1/3$ Laughlin state induces a charge of $-e/(3n)$. Therefore, the charge $-e/(3n)$ particle can be interpreted as a $\mathbb{Z}_n$ vortex in the $b$ sector, which is endowed with fractional charge due to its coupling to the $f$ sector.

$3n$ copies of the $\mathbb{Z}_n$ vortex, corresponding to the quasiparticle vector $(0,0,3n)$, gives the $f$ fermion, which is topologically equivalent to the electrically neutral composite boson $b$ when combined with an electron. This boson $b$, described by the quasiparticle vector $(0,-1,0)$, has $\mathbb{Z}_n$ fusion rules as $b^n$ is topologically trivial, and thus $b$ can be viewed as a $\mathbb{Z}_n$ charge. We see, then, that the topologically distinct classes of excitations form the group $\mathbb{Z}_{3n^2}$ under fusion. 

Another remark is in order here. The wave function $n\bar{n}111$ fully specifies the state of the $\mathbb{Z}_n$ superconductor, from which we have derived the $K$-matrix in Eq.~(\ref{eq:Kmatrix2_n_barn_111}). However, the converse is not true: demanding $\mathbb{Z}_n$ order does not fully specify the state. $\mathbb{Z}_n$ superconductors with several distinct topological structures are possible. The most conventional $\mathbb{Z}_n$ superconducting state of composite bosons would not have the second term on the right hand side of Eq.~(\ref{eq:Lagrangian_density_parton_nbarn111_composite_bosons}), $\frac{n}{4\pi} \beta d \beta$. The presence of this term indicates that the boson state also has a non-trivial topological response, which is possible because the $\mathbb{Z}_n$ gauge symmetry $b \rightarrow e^{2\pi i /n} b$ allows for the possibility that $b$ also forms a non-trivial $\mathbb{Z}_n$ symmetry-protected topological state. This is related to the fact that there are topologically distinct ways of gauging the $\mathbb{Z}_n$ subgroup of the $U(1)$ charge conservation symmetry, related to the third cohomology group $\mathcal{H}^3(\mathbb{Z}_n, U(1))$~\cite{Barkeshli14a}. We have not considered the other possibilities and their associated wave functions here.

\section{Edge theory}
\label{edgeSec}
The edge theory~\cite{Wen95,Moore98,Chang03} is determined by the above $K$ matrix
\begin{equation}
 \mathcal{L} ={1\over 4\pi} (K_{ij}\partial_t \phi_i \partial_x\phi_j - V_{ij} \partial_x \phi_i \partial_x\phi_j)
\end{equation}
in terms of three bosonic fields $\vec{\phi}=(\phi_{1},\phi_{2},\phi_{3})$ and the velocity matrix $V$. We can further add backscattering terms of the form
\begin{equation}
 \mathcal{\delta L} \propto \cos(\vec{\Lambda}^{\rm T}\cdot K \cdot \vec{\phi})
\end{equation}
where $\vec{\Lambda}$ is an integer valued vector. If $\vec{\Lambda}$ is a null vector, then the above backscattering term can gap out a pair of counterpropagating modes if the strength of the backscattering is large enough ~\cite{Haldane95,Levin13,Barkeshli13a} .
The $K$ matrix given in Eq.~(\ref{eq:Kmatrix2_n_barn_111}) has a null vector $\vec{\Lambda}^{\rm T}=(0,0,1)$ such that
\begin{equation}
\vec{\Lambda}^{\rm T}\cdot K \cdot \vec{\Lambda}=0. 
\end{equation}
This produces the scattering term 
\begin{equation}
 \mathcal{\delta L} \propto \cos(n\phi_{2}).
\end{equation}
Depending on the velocity matrix for the edge modes it is possible that the above backscattering term is irrelevant in the renormalization group (RG) sense in which case the edge theory with three gapless modes will be perturbatively stable. On the other hand, if the above backscattering term is either relevant in the RG sense or has a large amplitude it gaps out a pair of counter-propagating modes by pinning $\phi_{2}$ to a constant value. Generically, we expect this to be the case since there is no particular reason for the backscattering amplitude to be small. Once a pair of counter-propagating modes is gapped out the remaining chiral gapless mode is described by the usual $1/3$ Laughlin edge theory. \\

When $\phi_{2}$ is pinned to a constant value, it follows that $\langle e^{i\phi_{2}}\rangle \neq 0$. Physically this means that the $\mathbb{Z}_{n}$ gauge charge has condensed at the boundary, reducing the number of gapless modes to a single chiral mode. Note that in terms of the parton construction, the $\mathbb{Z}_n$ gauge charge corresponds to the boson $b$. 
Furthermore, note that the $\mathbb{Z}_n$ vortex carrying charge $-e/(3n)$ particle is created by the operator $e^{i\phi_{3}}$. Since~\cite{Wen95} 
\begin{equation*}
 [\partial_{x}\phi_{2}(x),\phi_{3}(y)] = 2\pi i [K^{-1}]_{2,3}\delta(x-y) = -i\pi \delta(x-y),
\end{equation*}
applying $e^{i\phi_{3}}$ creates a kink in $\phi_{2}$, which costs a finite energy. This implies that the charge $-e/(3n)$ particles cannot be created at arbitrary low energies at the edge. Physically, the condensation of the $\mathbb{Z}_{n}$ charge at the edge confines the $\mathbb{Z}_{n}$ vortex. For the same reason, it can be verified that all of the charge $(-e)k/(3n)$ particles with $k < n$ will be gapped at the edge. 

On the other hand, the charge $-e/3$ particle is created/annihilated by the operator $e^{i\phi_{1}}$ which commutes with $\phi_{2}$ and therefore the charge $-e/3$ excitations do not cause a kink in $\phi_{2}$. Consequently, the charge $-e/3$ particles can be created at arbitrary low energies at the edge. Any edge-based measurement of the fractional charge would therefore see $-e/3$ as the minimal quasiparticle at the lowest voltages and temperatures. A similar ``quasiparticle blocking'' effect for charge $-e/4$ non-Abelian quasiparticles was shown to be possible~\cite{Barkeshli15a} at even-denominator FQH states such as the Moore-Read Pfaffian~\cite{Moore91} when the edge is coupled to a superconductor. \\

If there is a separation of scales between the bulk energy gap $\Delta_{\rm bulk}$ of the system and the energy gap $\Delta_{\rm edge}$ corresponding to the pinning of $\phi_{2}$ in the edge theory, then at intermediate temperatures $\Delta_{\rm edge}<T<\Delta_{\rm bulk}$, or at intermediate voltages $\Delta_{\rm edge}<V<\Delta_{\rm bulk}$, the observed quasiparticle charge should be $-e/(3n)$. However, we do not know of a reason to generically expect such a separation of energy scales.

We note that in principle, there could be topologically distinct types of edge theories if the topological order contains distinct sets of bosons that can condense at the edge~\cite{Levin13,Barkeshli13a,Barkeshli15a}. Nevertheless, one can verify that the $\mathbb{Z}_{n}$ charge is the only topologically non-trivial boson. Therefore, there is only one edge phase with chiral central charge $c_{-}=1$. 

\section{The $n\bar{n}111$ state vis-{\'a}-vis experiments}
\label{sec:discussion}

We have introduced wave functions denoted by $n\bar{n}111$, which describe FQH states at filling factor $\nu=1/3$. We showed that the $n=2$ and $n=3$ members of this  sequence are good candidates to describe the $\nu=7/3$ FQHE. We have discussed how these states can be distinguished from the Laughlin 1/3 state. We now come to the current experimental status of our understanding of the 7/3 state and how the $n\bar{n}111$ states might relate to it.

The thermal Hall conductance $\kappa_{xy}$ has proved instrumental in distinguishing between different topological states at the same fraction~\cite{Banerjee17}. At temperatures much smaller than the gap and at length scales much longer than the equilibration length of the edges, $\kappa_{xy}$  takes a value proportional to the chiral central charge i.e., $\kappa_{xy}=c_{-}~[\pi^2 k_{\rm B}^2 /(3h)]T$~\cite{Kane97,Read00,Cappelli02}. The thermal Hall conductance at $7/3$ was measured to be close to $3~[\pi^2 k_{\rm B}^2 /(3h)]T$ (see Fig.~3a of Ref.~\cite{Banerjee17}) which indicates that the $1/3$ FQH state in the second Landau level state has $c_{-}\approx +1$. (The two filled lowest Landau levels of spin up and spin down contribute $2~[\pi^2 k_{\rm B}^2 /(3h)]T$ to $\kappa_{xy}$.) The $n\bar{n}111$  states as well as the Laughlin state are consistent with this value. We also note that the $n\bar{n}111$  states do not necessarily have a robust upstream neutral mode in their edge theories, since it can be gapped by appropriate backscattering terms.

The particle-hole conjugate of the $k=4$ Read-Rezayi state~\cite{Read99} also occurs at $\nu=1/3$ and has been put forth as a plausible candidate to describe the 7/3 FQHE~\cite{Friedman10,Peterson15}. This state has a chiral central charge of $-1$~\cite{Read99,Bishara08}. \emph{Assuming} full equilibration of edge modes and including the contributions of the two filled lowest Landau levels, the thermal Hall conductance of this state is predicted to be $\kappa_{xy}=1~[\pi^2 k_{\rm B}^2 /(3h)]T$, which is different from the experimentally measured value at $7/3$~\cite{Banerjee17}.

The Hall viscosity of an FQH state at $\nu=1/3$ is expected to be quantized~\cite{Read09}: $\eta_{H}=\hbar\mathcal{S}/(24\pi\ell^{2})$, where $\mathcal{S}$ is the shift. The $n\bar{n}111$  states as well as the Laughlin state~\cite{Laughlin83} occur at the same shift $\mathcal{S}=3$ and are therefore expected to carry the same Hall viscosity. We note here that the particle-hole conjugate of the $k=4$ Read-Rezayi state~\cite{Read99} has a shift $\mathcal{S}=-3$ and therefore has a different Hall viscosity than our  states. 

The topological entanglement entropy (TEE) of the $n\bar{n}111$ parton state is $\ln(\sqrt{3n^2})$. For the 7/3 state, the TEE has been computed from a density matrix renormalization group (DMRG) calculation and appears to converge to the value ascertained from the Laughlin state. However, the DMRG calculation of TEE shows strong system size dependence and thus it is difficult to reliably estimate its value in the thermodynamic limit. 

As we pointed out above, a remarkable difference between the Laughlin and the $n\bar{n}111$ states relates to the charge of the excitation. The charge of the elementary quasiparticle excitation of the $n\bar{n}111$ state is $-e/(3n)$. The experimental studies by Dolev \emph{et al.}~\cite{Dolev08,Dolev11} employing shot noise measurements found quasiparticles of charge only $-e/3$ at both $7/3$ and $8/3$. This is not inconsistent with the $n\bar{n}111$ state.  Even though this state has quasiparticles of charge $-e/(3n)$, as we explained above, the $-e/(3n)$ quasiparticle is ``gapped'' at the edge. Therefore, while we could add $-e/3$ particles at arbitrarily low energies at the edge, adding $-e/(3n)$ particles would cost a finite energy. Thus in shot noise measurements at arbitrarily small voltages, one would get $-e/3$ for the fractional charge even for the $n\bar{n}111$ state with $n\geq 2$. It is possible that at higher voltages or higher temperatures, the $-e/(3n)$ particle could also be excited, and the minimal charge could actually go from $-e/3$ to $-e/(3n)$ as the bias or temperature is increased. 

Venkatachalam \emph{et al.}~\cite{Venkatachalam11} have measured the local charge of the quasiparticles in the bulk at 7/3 using a scanning single electron transistor and found evidence for $-e/3$ charged quasiparticles. This favors the Laughlin state. However, a charge $-e/3$ could result in the $n\bar{n}111$ state with $n\geq 2$ if the elementary charge $-e/(3n)$ quasiparticles are bound in $n$-tuplets. Although we have not studied this issue quantitatively for the $n\bar{n}111$ state, calculations have indicated the quasiparticles of the Jain $n/(2n+1)$ states for $n>1$ generically have an attractive interaction~\cite{Lee02}. If that is the case, then, again, it is possible that raising the temperature or the bias could liberate the charge $-e/(3n)$ quasiparticles.

The ansatz ``$2\bar{2}111$'' implies the existence of multiple branches of magnetoroton excitations at 7/3, which arise from the different ways in which we can create particle-hole pairs in the $2$, $\bar{2}$ and $1$ sectors. There may be evidence of multiple roton minima in both exact diagonalization~\cite{Balram13b} and resonant inelastic light scattering experiments~\cite{Wurstbauer15}. 

Balram~\emph{et al.} have recently proposed\cite{Balram18a} for the SLL FQHE the sequence of states $\bar{n}\bar{2}111$ described by the wave functions $\Psi_{2n/(5n-2)} = \mathcal{P}_{\rm LLL} \Phi^*_{n}  \Phi^{*}_{2} \Phi^{3}_{1}$. In the spherical geometry these states have filling factor $\nu=2n/(5n-2)$ and shift $\mathcal{S}=(1-n)$. While $n=1$ produces a state that is essentially the same as $2/3$ Jain CF state~\cite{Balram16b}, $n=2$ and $3$ are plausible candidates for the second Landau level FQHE at $\nu=5/2$ and $2+6/13$, respectively~\cite{Balram18,Balram18a,Balram20}. We can analogously consider the states $n\bar{2}111$ described by the wave functions $\Psi_{2n/(5n+2)} = \mathcal{P}_{\rm LLL} \Phi_{n}  \Phi^{*}_{2} \Phi^{3}_{1}$. The $n=1$ case reproduces the Jain CF state at $2/7$~\cite{Jain89}; signatures of FQHE have been observed at $\nu=2+2/7=16/7$ and its particle-hole conjugate at $\nu=2+5/7=19/7$~\cite{Pan08}. The state $2\bar{2}111$ proposed in this work is the $n=2$ member of this sequence and thus fits in nicely with the other candidate states proposed for the SLL FQHE. 

Although we have only considered fully spin polarized states in this work, the parton theory admits the possibility of partially spin polarized and spin-singlet states at $1/3$. These are built from the partially spin polarized and spin-singlet versions of the $\nu=n$ IQH states. It is plausible that for some interaction, these unpolarized states have lower energy compared to the fully polarized one. The properties of these multi-component states remain to be explored. Intriguingly, a recent experiment using spin-resolved pulsed tunneling indicates that the $\nu=7/3$ state may not be fully spin polarized~\cite{Yoo19}. 

To the best of our knowledge our state provides the first example of a single component Abelian fractional quantum Hall state at $\nu=p/q$ (with $p,q$ coprime) which has (i) a torus ground state degeneracy that is greater than $q$, and (ii) quasiparticles carrying a charge of magnitude less than $e/q$. 

\begin{acknowledgments}
We acknowledge useful discussions with Hans Hansson, Mark S. Rudner, Steven H. Simon and Bo Yang. The Center for Quantum Devices is funded by the Danish National Research Foundation. This work was supported by the European Research Council (ERC) under the European Union Horizon 2020 Research and Innovation Programme, Grant Agreement No. 678862 and the Villum Foundation (A. C. B.). The work at Penn State was supported by the U. S. Department of Energy, Office of Basic Energy Sciences, under Grant no. DE-SC0005042 (J. K. J.). M. B. is supported by NSF CAREER (DMR-1753240), JQI-PFC-UMD and an Alfred P. Sloan Research Fellowship. Some of the numerical calculations were performed using the DiagHam package, for which we are grateful to its authors. Some portions of this research were conducted with Advanced CyberInfrastructure computational resources provided by The Institute for CyberScience at The Pennsylvania State University and the Nandadevi supercomputer, which is maintained and supported by the Institute of Mathematical Science’s High Performance Computing center. J. K. J. thanks the the Indian Institute Science, Bangalore, where part of this work was performed, for their hospitality, and the Infosys Foundation for
making the visit possible.
\end{acknowledgments}

\appendix

\section{Energies of various candidate states in graphene}
\label{app:energies_parton_Laughlin_LLL_GLL1}
In the main article we have considered 1/3 FQHE in the SLL of electrons that have a parabolic dispersion at zero magnetic field, as is the case in two-dimensional systems fabricated from semiconductor heterostructures and quantum wells. In this appendix we consider the competition between the $2\bar{2}111$, $3\bar{3}111$  and Laughlin states in the $n=0$ and $n=1$ LLs of Dirac electrons. This will be relevant to FQHE in graphene.

\begin{figure}[ht]
\begin{center}
\includegraphics[width=0.45\textwidth,height=0.3\textwidth]{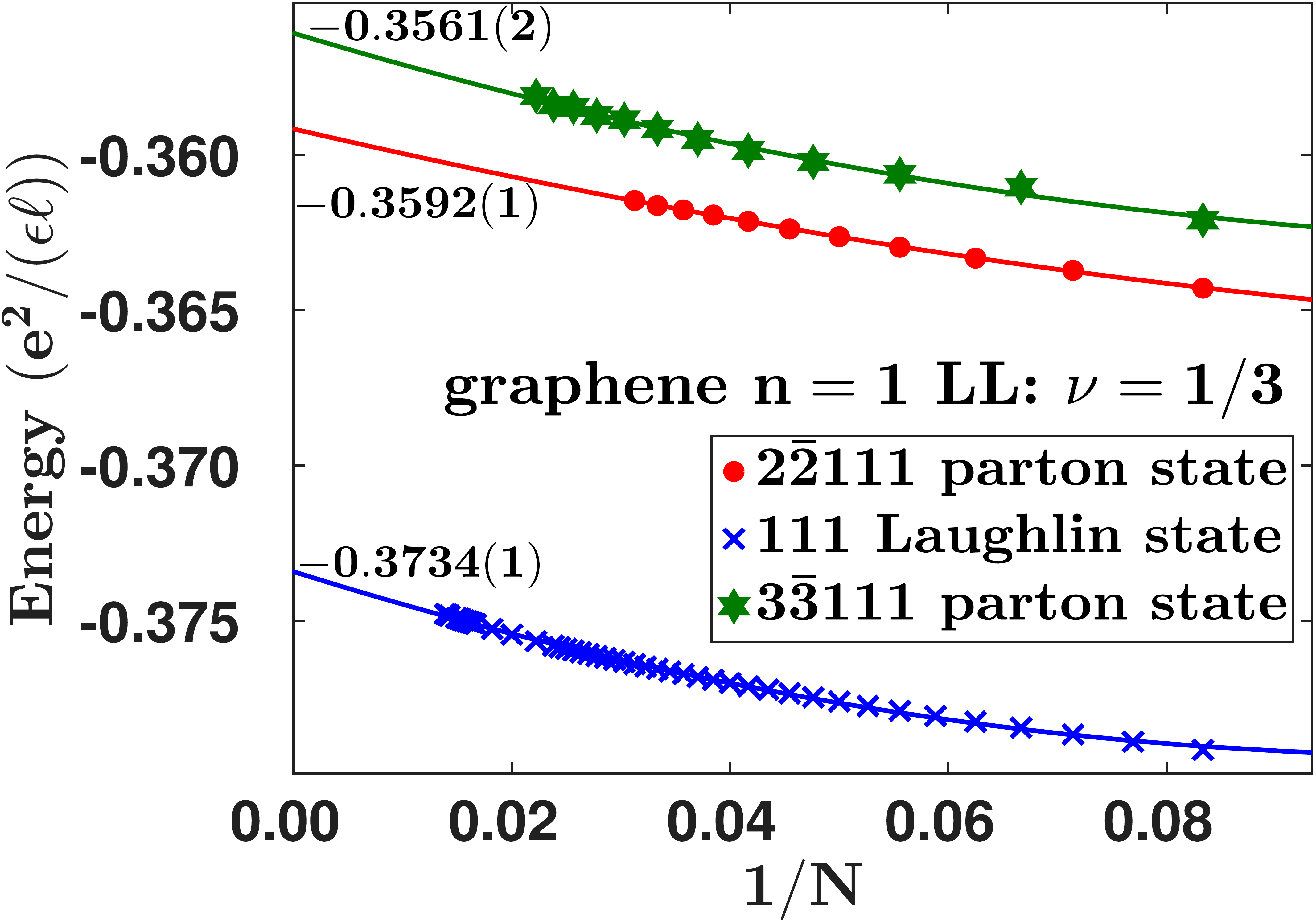} 
\caption{(color online) Same as in Fig.~\ref{fig:extrapolations_energies_7_3} but for $\nu=1/3$ in the $n=1$ LL of Dirac electrons in graphene.
}
\label{fig:extrapolations_energies_1_3_other_LLs}
\end{center}
\end{figure}

The physics of FQHE in the $n=0$ LLs of graphene and GaAs are identical~\cite{Nomura06,Goerbig06,Apalkov06,Toke06} to the extent finite width and LL mixing effects can be neglected. The comparison shown in Fig.~\ref{fig:extrapolations_energies_1_3} therefore applies to the $n=0$ graphene LL as well. Fig.~\ref{fig:extrapolations_energies_1_3_other_LLs} shows the Coulomb ground state energies of these candidate states at $\nu=1/3$ in the $n=1$ LL of graphene. We simulate the the $n=1$ LL of graphene using the effective interaction in the LLL given in Ref.~\cite{Balram15c}. All numbers quoted in Fig.~\ref{fig:extrapolations_energies_1_3_other_LLs} include density correction~\cite{Morf86} and also the electron-background and background-background interaction. We find that, in contrast to the $n=1$ LL of semiconductor based systems, the $2\bar{2}111$ and $3\bar{3}111$ states are not competitive in the $n=1$ graphene LL. This is consistent with the expectation that the description in terms of weakly interacting composite fermions is valid in both the $n=0$ and $n=1$ LLs of graphene~\cite{Amet15,Balram15c,Zeng18,Kusmierz18}. 

\section{$n\bar{n}$  is a $\mathbb{Z}_{n}$ superconductor}
\label{app:effective_field_theory_parton_nbarn}

In this Appendix and the next, we obtain the field theoretical description of the $n\bar{n}111$ state. To this end let us begin with the $n\bar{n}$ state with the wave function
\begin{equation}
 \Psi^{n\bar{n}} = \Phi_{n}\Phi_{\bar{n}},
 \label{eq:parton_n_barn}
\end{equation}
This wave function occurs at zero magnetic field, which can be seen by noting that it is real. This wave function is symmetric under the exchange of the coordinates of any two particles and therefore describes a many-body state of bosons. The effective field theory predicts that this wave function describes a state with off-diagonal long-range order (ODLRO), where the $U(1)$ global charge conservation symmetry is broken down to a $\mathbb{Z}_{n}$ subgroup~\cite{Barkeshli13}. The $n=1$ case corresponds to the wave function of a superfluid of bosons and has been used to describe a possible continuous phase transition between a Bose superfluid and the $\nu=1/2$ Laughlin state~\cite{Barkeshli14,Shapourian16}. 

To understand the properties of the  wave function given in Eq.~(\ref{eq:parton_n_barn}) we consider the associated parton decomposition
\begin{equation}
 b = f_{1} f_{2},
 \label{eq:bosonic_parton_n_barn}
\end{equation}
where $f_{1}$ and $f_{2}$ form $\nu=n$ and $\nu=-n$ IQH states, respectively. This ansatz has an emergent $U(1)$ gauge field $\alpha$ associated with the transformation $f_{1}\rightarrow e^{i\theta}f_{1}$ and $f_{2}\rightarrow e^{-i\theta}f_{2}$ which leaves $b$ invariant, and under which $\alpha_{\mu}\rightarrow \alpha_{\mu} +\partial_{\mu}\theta$, . The boson $b$ carries a unit charge under the background external $U(1)$ electromagnetic gauge field $A$. In the parton decomposition we assign charges $1$ and $0$ to $f_{1}$ and $f_{2}$ respectively. [In the most general setting, we can assign charges $q_1$ to $f_1$ and $q_2$ to $f_2$ with the constraint $q_1 + q_2 = -e$, where $-e$ is the electronic charge. These different choices of the charges of the partons can be related to each other by a shift of the emergent $U(1)$ gauge field $\alpha \rightarrow \alpha + q_2A$, although care must be taken to ensure global consistency conditions related to compactness of the gauge fields are properly respected. Choosing integer charges for all partons ensures that these consistency conditions are properly taken into account.]  

The effective field theory obtained by integrating out the parton fields is described by the Lagrangian density (for convenience we set $\hbar=c=e=1$): 
\begin{equation}
  \label{ZnSPT}
 \mathcal{L} = \frac{n}{4\pi}(\alpha+A) d (\alpha+A) - \frac{n}{4\pi} \alpha d \alpha - \frac{1}{g^{2}} (\epsilon_{\mu\nu\lambda}\partial_{\nu} \alpha_{\lambda})^{2},
\end{equation}
where we have used the shorthand notation $\alpha d \alpha \equiv \epsilon^{\mu\nu\lambda} \alpha_{\mu}\partial_{\nu} \alpha_{\lambda}$. In the above Lagrangian density the first term corresponds to integrating out $f_{1}$ ($n$ modes propagating in the forward direction and carrying charge $1$ with respect to both the emergent gauge field $\alpha$ and the external gauge field $A$), the second term corresponds to integrating out $f_{2}$ ($n$ modes counter-propagating and carrying charge $-1$ with respect to $\alpha$ and $0$ charge with respect to the external gauge field $A$). The last term is the Maxwell action with the coupling constant $1/g^{2}$ for the internal gauge field $\alpha$. Usually the Maxwell term for $\alpha$ is ignored, but we must include it here since the leading order Chern-Simons (CS) term for $\alpha$ cancels out resulting in the Lagrangian density given by
\begin{equation}
 \mathcal{L} = \frac{n}{2\pi}A d \alpha + \frac{n}{4\pi} A d A - \frac{1}{g^{2}} (\epsilon_{\mu\nu\lambda}\partial_{\nu} \alpha_{\lambda})^{2}.
\label{eq:Lagrangian_density_parton_n_barn_appendix}
\end{equation}

We will now show that the Lagrangian density given in Eq.~(\ref{eq:Lagrangian_density_parton_n_barn_appendix}) describes a superconducting state where the $U(1)$ global symmetry has been broken down to $\mathbb{Z}_{n}$. The internal gauge field $\alpha$ is gapless since the effective action is just the Maxwell action. In $(2+1)D$, a gapless $U(1)$ gauge field is dual to a real scalar field, i.e., a Goldstone mode. Since the conserved boson current is
\begin{equation}
 j_b^{\mu} = \frac{\delta \mathcal{L}}{\delta A_{\mu}} = \frac{n}{2\pi}\epsilon^{\mu\nu\lambda}\partial_{\nu}[\alpha_{\lambda}+A_{\lambda}],
\end{equation}
fluctuations in $\alpha$ lead to current and density fluctuations. \\

In this theory the vortices of the superconductor correspond to the gapped fermionic excitations of the parton states as these are minimally coupled to the dual Goldstone mode $\alpha$. To see that this theory describes a charge $n$ superconductor we let the background gauge field $A$ be dynamical. Then the dual Goldstone mode $\alpha$ should be gapped by the Anderson-Higgs mechanism. The low-energy theory is
\begin{equation}
 \mathcal{L} = \frac{n}{2\pi}\alpha d A + \alpha_\mu (j^\mu_{f_1}-j^\mu_{f_2})+ \cdots,
\label{eq:Lagrangian_density_parton_n_barn_AH_mechanism_appendix}
\end{equation}
where $j_{f_1}$ and $j_{f_2}$ are the currents associated with the partons $f_{1}$ and $f_{2}$ and the $\cdots$ indicate higher derivative terms. The equation of motion for $\alpha$ is the constraint:
\begin{equation}
2\pi B^{\mu} = \epsilon^{\mu\nu\lambda}\partial_{\nu}A_{\lambda} = \frac{2\pi}{n} (j^{\mu}_{f_2}-j^{\mu}_{f_1}),
\end{equation}
where $B=\nabla \times A$ is the external magnetic field. This shows that the gapped excitations associated with $f_{1}$ and $f_{2}$ each carry $2\pi/n$ units of flux, as expected for the vortices of a charge $n$ superconductor. \\

Another way to see this is to introduce a field $\xi^{\mu}=(1/2\pi)\epsilon^{\mu\nu\lambda}\partial_{\nu} \alpha_{\lambda}$ in Eq.~(\ref{eq:Lagrangian_density_parton_n_barn_appendix}) and a Lagrange multiplier $\varphi$ to enforce $\partial_{\mu} \xi^{\mu}=0$. Integrating out $\xi^{\mu}$ then gives
\begin{equation}
 \mathcal{L} \propto (\partial\varphi - n A)^{2},
\label{eq:Lagrangian_density_charge_n_SC_appendix}
\end{equation}
which is the Lagrangian of a charge $n$ superconductor.

We note that the effective action of Eq.~(\ref{ZnSPT}) has a term $\frac{n}{4\pi} \epsilon^{\mu \nu \lambda} A_\mu \partial_\nu A_\lambda$, which is not present in the usual effective action for a $\mathbb{Z}_n$ superconductor. This is a topological response term whose origin can be traced to the charge assignment of the partons. While it is not clear that this term is meaningful for the superconducting state of $b$, which is gapless due to the Goldstone mode dual to the $U(1)$ gauge field $\alpha$, this term does have important consequences for the fractional statistics of the charge $-e/(3n)$ quasiparticles of the $n\bar{n}111$ state.

We note that one can derive an alternate effective field theory by representing each filled Landau level with an independent $U(1)_1$ Chern-Simons gauge field, which leads to the action:
\begin{align}
  \nonumber \\
  \mathcal{L} = & -\frac{1}{4\pi} \sum_{I=1}^n a^I d a^I + \frac{1}{4\pi} \sum_{I=n+1}^{2n} a^I d a^I 
  \nonumber \\
&+  \frac{1}{2\pi} \sum_{I=1}^n \alpha d (a^I - a^{n+I}) + \frac{1}{2\pi} \sum_{I=1}^n A d a^I
\label{eq:alternate_Lagrangian_density_parton_n_barn_bosons_appendix}
\end{align}
The boson current in this setup is then given by
\begin{align}
 j_\mu^{(b)} = \frac{1}{2\pi} \epsilon^{\mu\nu\lambda} \sum_{I=1}^n \partial_\nu a_\lambda^I.
  \end{align}
Integrating out the $a^I$ then gives back the Lagrangian density described above. 

\section{$n\bar{n}111$  wave functions}
\label{app:effective_field_theory_parton_nbarn111}

Let us now consider the following wave function
\begin{equation}
 \Psi^{n\bar{n}111} = \Phi_{n}\Phi_{\bar{n}}\Phi^{3}_{1}.
 \label{eq:parton_n_barn_111_appendix}
\end{equation}
The associated parton decomposition is
\begin{equation}
 \wp = b f,
 \label{eq:bosonic_parton_n_barn_appendix}
\end{equation}
where $b$ forms the $n\bar{n}$ state and $f$ forms a $\nu=1/3$ Laughlin state~\cite{Laughlin83}. As discussed in the main text, $b$ can be interpreted as the composite boson in the framework of Ref. \onlinecite{Zhang89}. The above ansatz has an emergent $U(1)$ gauge field $\beta$ associated with the transformation $b\rightarrow e^{i\Theta}b$ and $f\rightarrow e^{-i\Theta}f$ which leaves $\wp$ invariant, and under which $\beta_{\mu}\rightarrow \beta_{\mu} +\partial_{\mu}\Theta$. As we have seen in the previous section $n\bar{n}$ corresponds to a state that breaks the $U(1)$ symmetry associated with $b$ down to $\mathbb{Z}_{n}$. This implies that $n\bar{n}111$ state can be thought of as stacking a $\mathbb{Z}_{n}$ symmetric state on top of the $1/3$ Laughlin state, where the $\mathbb{Z}_{n}$ symmetry is a subgroup of the $U(1)$ global symmetry of the Laughlin state. The projection of $n\bar{n}$ to $111$ then corresponds to gauging this $\mathbb{Z}_{n}$ symmetry. In other words, the $n\bar{n}111$ state is related to the $1/3$ Laughlin state by gauging a $\mathbb{Z}_{n}$ subgroup of the global $U(1)$ symmetry. A general theory of gauging discrete symmetries in topologically ordered systems was presented in Ref.~\cite{Barkeshli14a}. Alternatively, since $b$ is the composite boson, this state can be understood as a state where the composite boson forms a $\mathbb{Z}_n$ superconductor.   \\

Taking $f$ to have electric charge $1$ and $b$ to have electric charge 0 under the external electromagnetic gauge field, the low-energy effective field theory of the $n\bar{n}111$ parton state is given by
\begin{equation}
 \mathcal{L} = -\frac{3}{4\pi}\tilde{a} d \tilde{a} + \frac{1}{2\pi} (\beta + A) d \tilde{a}+\frac{n}{2\pi} \beta d \alpha + \frac{n}{4\pi} \beta d\beta,
\label{eq:Lagrangian_density_parton_n_barn_111_appendix}
\end{equation}
where $\tilde{a}$ and $\alpha$ describe the $\nu=1/3$ Laughlin state $f$ and the bosonic state $b$ defined in Eq.~(\ref{eq:bosonic_parton_n_barn_appendix}) respectively. The second term on the right hand side of the above equation describes the coupling of the fermionic Laughlin sector $f$ to the emergent gauge field $\beta$ and to the external gauge field $A$. The last two terms on the right hand side of the above equation correspond to the effective theory of the $n\bar{n}$ state. They are obtained from Eq.~(\ref{eq:Lagrangian_density_parton_n_barn_appendix}) by replacing the external electromagnetic field $A$ by the emergent gauge field $\beta$. The effective field theory is thus a $U(1)^{3}$ Chern-Simons theory. Let us define a new set of gauge fields $(a^{1},a^{2},a^{3})=(\tilde{a},\beta,\alpha)$ to write the Lagrangian density in the form~\cite{Wen95}
\begin{equation}
 \mathcal{L} = -\frac{1}{4\pi}K^{(f)}_{\rm IJ}a^{\rm I}\partial a^{\rm J},
\end{equation}
where $K^{(f)}$is given by
\begin{equation}
K^{(f)} =  
\begin{pmatrix} 
      3 & -1 & 0 \\
      -1 & -n & -n \\
      0 & -n  & 0 \\
   \end{pmatrix}.
\label{eq:Kmatrix_n_barn_111_appendix}   
\end{equation}
Since we take $f$ and $b$ to have electric charge $1$ and $0$ respectively, we can take the corresponding charge vector of the CS theory to be $\vec{t}=(1,0,0)^{\rm T}$.

$K^{(f)}$  has two positive and one negative eigenvalues which indicates that the $n\bar{n}111$ state hosts two forward propagating and one backward moving mode resulting in a chiral central charge of $c_{-}=+1$. The ground state degeneracy on a manifold of genus $g$ is $|{\rm Det}(K^{(f)})|^{g}=(3n^{2})^{g}$. In particular, the ground state degeneracy on a torus is $3n^{2}$. This can be understood by noting that on a torus, we could pick any $\mathbb{Z}_n$ valued flux going through the two cycles of the torus, which, in addition to the factor of $3$ coming from the Laughlin $1/3$ state, gives a factor of $n^2$ to the ground state degeneracy.

Furthermore, observe that
\begin{align}
[K^{(f)}]^{-1} = \frac{1}{3n^2}\left(\begin{matrix}
n^2 & 0 & -n \\
0 & 0 & -3n \\
-n & -3n & 3n+1 \\
\end{matrix} \right).
\end{align}
The topologically non-trivial quasiparticles can be labeled by an integer vector $\vec{l}$, which indicates their gauge charges under the above gauge fields. $\vec{l}$ corresponds to a topologically trivial particle if and only if $K^{-1} \vec{l}$ is an integer vector. Therefore we can see that the quasiparticles labeled $\vec{l}_a^T = (0,0,a)$ are topologically non-trivial unless $a$ is a multiple of $3n^2$. This shows that the structure of the quasiparticle fusion rules is that of a $\mathbb{Z}_{3n^2}$ group. 

An intuitive way of understanding why there are $3n^2$ distinct quasiparticle sectors is as follows. First, $n$ charge $-e/(3n)$ quasiparticles make a single charge $-e/3$ quasiparticle. Three of these make what we have called an $f$-fermion, which is topologically equivalent to a single composite boson $b$. A collection of $n$ such composite bosons becomes a part of the condensate, and is thus topologically trivial. Altogether, that gives $3n^2$ topologically distinct quasiparticle sectors, which form a $\mathbb{Z}_{3n^2}$ group under fusion. 

However note that the particles / holes in the filled Landau level states of the $\Phi_n \Phi_{\bar{n}}$ factor are fermions, whose statistics is then transmuted due to the coupling to the Chern-Simons gauge fields. This means that the usual formula for the quasiparticle statistics~\cite{Wen95} $\theta_{\vec{l}} = \pi \vec{l}^T [K^{(f)}]^{-1} \vec{l}$ will only be correct modulo $\pi$, unless we properly take into account the fact that the quasiparticles can be fermions before being dressed by the CS gauge fields. For this reason, we have included the superscript $(f)$ in $K^{(f)}$. To take this into account properly, we can derive an alternate effective field theory as follows.

We use Eq.~(\ref{eq:alternate_Lagrangian_density_parton_n_barn_bosons_appendix}) for the effective action of the bosons, to obtain the following effective action for the $n\bar{n} 111$ state:
\begin{widetext}
\begin{align}
\mathcal{L} =& -\frac{1}{4\pi} \sum_{I=1}^{n} a^I d a^I + \frac{1}{4\pi} \sum_{I=n+1}^{2n} a^I d a^I + \frac{1}{2\pi} \sum_{I=1}^n \alpha d (a^I - a^{n+I} )
  - \frac{3}{4\pi} \tilde{a} d \tilde{a} + \frac{1}{2\pi} \sum_{I=1}^n \beta d (a^I - \tilde{a}) + \frac{1}{2\pi} A d \tilde{a}
\nonumber \\
&+ \frac{1}{2\pi} \sum_{I=1}^n ( I - \frac{1}{2}) a^I d \omega - \frac{1}{2\pi} \sum_{I=n+1}^{2n} ( I - \frac{1}{2}) a^I d \omega + \frac{3}{2} \frac{1}{2\pi} \tilde{a} d \omega,
\end{align}
\end{widetext}
where we have included the coupling to the spin connection $\omega_\mu$ to obtain the spin vector and shift of the resulting state. This has $2n + 3$ dynamical gauge fields. If we integrate out $a^I$, which corresponds to integrating out the fluctuations associated with the filled parton Landau levels, then we obtain the effective action described above. Alternatively, we can integrate out the constraint gauge fields $\alpha$ and $\beta$, which enforce the constraints $j_{f_1} = j_{f_2}$ and $j^{(b)} = j^{(f)}$, respectively. That is, integrating out $\alpha$ and $\beta$ enforces:
  \begin{align}
    \sum_{I=1}^n a^I &= \sum_{I=1}^n a^{n+I},
    \nonumber \\
    \tilde{a} &= \sum_{I=1}^n a^I . 
  \end{align}
  Now we have $2n-1$ gauge fields, described by the following action:
\begin{widetext}
\begin{align}
  \mathcal{L} = & -\frac{1}{4\pi} \sum_{I=1}^{n} a^I d a^I + \frac{1}{4\pi} \sum_{I=n+1}^{2n-1} a^I d a^I + \frac{1}{4\pi} (\sum_{I=1}^n a^I -\sum_{I=n+1}^{2n-1} a^I) d  (\sum_{I=1}^n a^I - \sum_{I=n+1}^{2n-1} a^I)
                  - \frac{3}{4\pi} \sum_{I,J = 1}^n a^I d a^J
                  \nonumber \\
                  &+ \frac{1}{2\pi} \sum_{I=1}^n  A d a^I + \frac{3}{2} \frac{1}{2\pi}  \sum_{I=1}^n a^I d \omega
                  \nonumber \\
  =& \frac{1}{4\pi} \sum_{I,J = 1; I \neq J}^n a^I d a^J - \frac{3}{4\pi} \sum_{I,J=1}^n a^I d a^J + \frac{1}{4\pi} \sum_{I,J = n+1}^{2n-1} (1 + \delta_{IJ}) a^I d a^J -
     \frac{1}{2\pi} \sum_{I=1}^n \sum_{J = n+1}^{2n-1} a^I d a^J
     \nonumber \\
    & + \frac{1}{2\pi} \sum_{I=1}^n A d a^I  + \frac{3}{2} \frac{1}{2\pi}  \sum_{I=1}^n a^I d \omega
\end{align}
\end{widetext}

We can summarize this $K$ matrix as follows:
\begin{align}
  \label{Kmatrix}
  K_{IJ} = \left\{
  \begin{array}{ll}
     3 - 1 + \delta_{IJ}& I,J \leq n\\
    -1 - \delta_{IJ} & I, J > n \\
      1 & \text{ otherwise}\\
    \end{array} \right.
\end{align}
with the charge vector
\begin{align}
  t_I = \left\{ \begin{array}{ll}
                  1 & I \leq n \\
                  0 & \text{ otherwise} \\
                  \end{array} \right.
  \end{align}
  and spin vector
\begin{align}
\mathfrak{s}_I = 3/2  \left\{ \begin{array}{ll}
                  1 & I \leq n \\
                  0 & \text{ otherwise} \\
                  \end{array} \right.
\end{align}
  
For $n = 2$, we therefore have the $K$ matrix:
\begin{align}
  K = \left( \begin{matrix}
3 & 2 & 1 \\
2 & 3 & 1 \\
1 & 1  & -2\\
      \end{matrix} \right)
  \end{align}
  with charge vector $\vec{t} = (1,1,0)^{\rm T}$ and spin vector $\vec{\mathfrak{s}} = 3/2(1,1,0)^{\rm T}$. We see that $|{\rm Det}(K)| = 12$. For $n = 3$, we have the $K$ matrix
\begin{align}
  K = \left(\begin{matrix}
      3 & 2 & 2 & 1 & 1 \\
      2 & 3 & 2 & 1 & 1 \\
      2 & 2& 3 & 1 & 1\\
      1 & 1 & 1 &  -2 & -1 \\
      1 & 1 & 1 & -1 & -2 \\
      \end{matrix} \right)
\end{align}

One can check that the inverse is given by
\begin{align}
(K^{-1})_{IJ} = \frac{1}{3n^2}\left\{
  \begin{array}{ll}
    1 - 3n + 3n^2 \delta_{IJ} & I,J \leq n\\
    1 + 3n - 3n^2 \delta_{IJ} & I, J > n \\
    1 & \text{ otherwise}
    \end{array} \right.
\end{align}

The shift is given by~\cite{Wen95}
\begin{align}
\mathcal{S} = \frac{2 \vec{t}^T \cdot K^{-1} \cdot \vec{\mathfrak{s}}}{\vec{t}^T \cdot K^{-1} \cdot \vec{t}} = 3 .
  \end{align}

A quasiparticle labeled by the vector $\vec{l}$ is topologically trivial if $K^{-1} \vec{l}$ is an integer vector. To see the fusion rule structure, let us consider the quasiparticles described by the vectors $l_{a;I} = a\delta_{I1}$. We can see that $\vec{l}_a$ is topologically non-trivial for all $a = 1,\cdots, 3n^2-1$. This implies that the quasiparticle $\vec{l}_1$ has $\mathbb{Z}_{3n^2}$ fusion rules. Furthermore, since $|{\rm Det}(K)| = 3n^2$, we see that all of the quasiparticles are described by $\vec{l}_a$, with exchange statistics and charge~\cite{Wen95}
\begin{align}
  \theta_a &= \pi \vec{l}_a \cdot K^{-1} \cdot \vec{l}_a  = \frac{a^2 \pi (3n(n-1) + 1)}{3n^2}
  \nonumber \\
  \mathcal{Q}_a &= (-e)\sum_{I=1}^n K^{-1}_{I1} = \frac{a(-e)}{3n} . 
  \end{align}

Finally, we observe that all of these topological properties can be reproduced using a simple $3 \times 3$ K matrix
 \begin{align}
\label{Kmatrix2}
K = \left(\begin{matrix}
3 & -1 & 0 \\
-1 & -n(n-1) & -n \\
0 & -n & 0 \\
\end{matrix} \right).
 \end{align}
 with charge vector $\vec{t}= (1,0,0)^{\rm T}$. The difference between this $K$-matrix and $K^{(f)}$ derived above in Eq.~(\ref{eq:Kmatrix_n_barn_111_appendix}) is the replacement of $-n$ with $-n(n-1)$ in the center $(2,2)$ entry, which properly gives the full statistics (modulo $2\pi$) of the quasiparticles. In fact, the $(2,2)$ entry being non-zero implies that we should think of the composite boson not as forming a topologically trivial $\mathbb{Z}_n$ superconductor, but rather a non-trivial $\mathbb{Z}_n$ symmetry-protected topological (SPT) state. For reference, the inverse of the $K$-matrix defined in Eq.~(\ref{Kmatrix2}) is
\begin{align}
K^{-1} = \frac{1}{3n^2}\left(\begin{matrix}
n^2 & 0 & -n \\
0 & 0 & -3n \\
-n & -3n & 1 + 3n(n-1) \\
\end{matrix} \right).
  \end{align} 

The fact that the two $K$-matrices of Eqs.~(\ref{Kmatrix2}) and (\ref{Kmatrix}) reproduce the same topological order implies that there should be an $SL(2n-1;\mathbb{Z})$ transformation $W$ [$W$ is a $(2n-1) \times (2n-1)$ integer matrix with unit determinant] which takes the $K$ matrix of Eq.~(\ref{Kmatrix}) to that of Eq.~(\ref{Kmatrix2}) up to a direct sum of $n-2$ trivial $\sigma^z$ factors, i.e., 
\begin{equation}
W K_1 W^T = K_2 \oplus \underbrace{\sigma_z \oplus \sigma_z \cdots \oplus \sigma_z}_{\text{$n-2$}}. 
\label{eq:W_transform_relate_Ks}
\end{equation}
Here $K_{1}$ is the $K$ matrix of Eq.~(\ref{Kmatrix}) and $K_{2}$ is the $K$ matrix of Eq.~(\ref{Kmatrix2}). For $n=1$, $K_{1}$ and $K_{2}$ swap places in Eq.~(\ref{eq:W_transform_relate_Ks}). The additional factors of $\sigma_z$ on the right hand side of Eq.~(\ref{eq:W_transform_relate_Ks}) can be interpreted as adding pairs of trivial counterpropagating modes.

\subsection{$n=2$ case}
  
In this subsection we specifically focus on the $n=2$ case. We have $12$ quasiparticles (equal to the ground state degeneracy on the torus). The fusion rules break up into a $\mathbb{Z}_{12}$ structure which, using the $K$-matrix of Eq.~(\ref{Kmatrix2}) we can label by the integer vectors
\begin{equation}
\vec{l}_{a} = (0,0,a)^{\rm T},~a=0,1,2,\cdots, 11
\label{eq:vector_n_barn_111}   
\end{equation}
These have electric charges~\cite{Wen95}
\begin{equation}
\mathcal{Q}_{a} = (-e)\vec{t}^{\rm T}\cdot K^{-1} \cdot \vec{l}_{a} =ae/6
\label{eq:charge_n_barn_111}   
\end{equation}
and exchange statistics~\cite{Wen95}
\begin{equation}
\theta_{a} = \pi~\vec{l}_{a}^{\rm T}\cdot K^{-1} \cdot \vec{l}_{a} = \frac{7 \pi a^2}{12} 
\label{eq:statistics_n_barn_111}   
\end{equation}
Therefore the minimal electric charge is $1/6$ the electron charge. Physically we can think of this as a $\pi$ flux of the $\beta$ gauge field, which induces charge $-e/6$ in the Lauglin sector.

The quasiparticle corresponding to the vector $\vec{l}= (0,0,6)^{\rm T}$ has exchange statistics $\theta_{\vec{l}}(\text{mod}~2\pi)= \pi$ and carries unit electric charge $\mathcal{Q}_{6}=e$. The quasiparticle corresponding to $\vec{l}= (0,-1,-6)^{\rm T}$ is topologically trivial, has statistics $\theta_{\vec{l}}(\text{mod}~2\pi)= \pi$, and electric charge $\mathcal{Q}_{\vec{l}} = -e$. It can thus be identified with the electron. Fusing the electron with the quasiparticle $\vec{l}= (0,0,6)^{\rm T}$ therefore gives a topologically non-trivial, electrically neutral bosonic excitation, described by the integer vector $(0,-1,0)$. We see that in a concrete sense the $2\bar{2}111$ state has a hidden $\mathbb{Z}_{2}$ topological order. The topologically non-trivial neutral bosonic excitation is the $\mathbb{Z}_{2}$ charge, while the charge $-e/6$ quasiparticle is the $\mathbb{Z}_{2}$ vortex. From Eq.~(\ref{eq:statistics_n_barn_111}) we can see that the statistics of the charge $-e/6$ $\mathbb{Z}_2$ vortex, which is described by the vector $\vec{l}_{-1}= (0,0,-1)^{\rm T}$ is $\theta_{-1} = 7 \pi/12 = \pi/12 + \pi/2 $ [Note that $\vec{l}_{-1}$ is topologically equivalent to $\vec{l}_{11}$ given in Eq.~(\ref{eq:vector_n_barn_111}). Addition of two electrons to the quasiparticle described by $\vec{l}_{11}$ results in the $-e/6$ quasiparticle described by $\vec{l}_{-1}$.]. The $\pi/12$ factor can be understood from the fact that the quasiparticle has flux $\pi$ and charge $-e/6$, while the $\pi/2$ factor can be understood from the fact that the composite boson sector is forming a $\mathbb{Z}_2$ SPT state as mentioned above. 

\bibliography{biblio_fqhe}
\bibliographystyle{apsrev}

\end{document}